\documentclass[showpacs, pra,twocolumn,preprintnumbers ,amsmath, amssymb, superscriptaddress, aps]{revtex4-2}
\usepackage{color}
\usepackage{amsmath,amssymb}
\usepackage{pifont}% http://ctan.org/pkg/pifont
\usepackage{amssymb}  % More symbols
\usepackage{bbold}
\usepackage{float}
\usepackage{subfloat}
\usepackage[caption=false]{subfig}
\usepackage{tikz}
\usepackage{makecell}
\usepackage{subfig}
\usepackage{pifont}   % Ding symbols
\usepackage{graphicx} % Include figure files
\graphicspath{{Figures/}}
\usepackage{dcolumn}  % Align table columns on decimal point
\usepackage{bm}       % bold math
\usepackage{multirow} % Table functions
\usepackage[colorlinks]{hyperref}
\usepackage{placeins}% For making floats not move around everywhere
\usepackage{mathtools}
\usepackage{appendix}

\newcommand{\tr}{\mathrm{tr}}

\begin{document}
	\title{Transport properties through alternating borophene and graphene superlattices}
	%\title{Transport properties in monolayer borophene-graphene junctions superlattices}
	\date{\today}
	\author{Nadia Benlakhouy}
	\email{benlakhouy.a@ucd.ac.ma}
	\affiliation{Laboratory of Theoretical Physics, Faculty of Sciences, Choua\"ib Doukkali University, PO Box 20, 24000 El Jadida, Morocco}
	\author{Abderrahim El Mouhafid}
	\email{elmouhafid.a@ucd.ac.ma}
	\affiliation{Laboratory of Theoretical Physics, Faculty of Sciences, Choua\"ib Doukkali University, PO Box 20, 24000 El Jadida, Morocco}
	\author{Ahmed Jellal}
	\email{a.jellal@ucd.ac.ma}
	\affiliation{Laboratory of Theoretical Physics, Faculty of Sciences, Choua\"ib Doukkali University, PO Box 20, 24000 El Jadida, Morocco}
	%\affiliation{Saudi Center for Theoretical Physics, Dhahran, Saudi
		%Arabia}
	\affiliation{Canadian Quantum Research Center, 204-3002 32 Ave Vernon,  Vernon BC V1T 2L7, Canada}

	\pacs{78.67.Wj, 05.40.-a, 05.60.-k, 72.80.Vp\\
		{\sc Keywords}: Graphene, borophene, periodic barrier, junctions, transmission, conductance, Fano factor.}

	\begin{abstract}
		
			The electronic transport properties of two junctions (BGB, GBG) made of borophene (B) and graphene (G) are investigated. Using the transfer matrix method with Chebyshev polynomials, we have studied single and multiple barriers in a superlattice configuration.
			 We showed that a single barrier exhibits remarkable tilted transport properties, with perfect transmission observed for both junctions under normal incidence. We found that robust superlattice transmission is maintained for multiple barriers, particularly in the BGB junction. It turns out that by varying the incident energy, many gaps appear in the transmission probability. The number, width, and position of these transmission gaps can be manipulated by adjusting the number of cells, incident angle, and barrier characteristics. For diffuse transport, we observed  considerable variations in transmission probability, conductance and the Fano factor, highlighting the sensitivity of these junctions to the physical parameters. We showed different behaviors between BGB and GBG junctions, particularly with respect to the response of conductance and Fano factor when barrier height varies. For ballistic transport, we have seen that the minimum {scaled conductance} is related to the maximum Fano factor, demonstrating their control under specific conditions of the physical parameters. Analysis of the length ratio (geometric factor) revealed some remarkable patterns, where {scaled conductance} and the Fano factor converged to certain values as the ratio approached infinity.

	\end{abstract}

	\maketitle
	\section{Introduction}
	%%%%%%%%%%%%%%%%%%%%%
	The discovery of graphene, a single layer  of carbon atoms arranged in a hexagonal lattice, was a milestone in materials science and earned the Nobel Prize in 2010 \cite{novoselov2004electric, Novoselov2005two}. The two-dimensional (2D) wonder material demonstrated superior electrical, mechanical, and thermal properties, laying the foundations for a wide range of applications \cite{neto2009electronic,silvestre2015folded, zhao2015graphene, ji2012atomic}. Following in the footsteps of graphene, borophene, a monolayer of boron atoms with unique honeycomb structures, has emerged as a promising nanomaterial \cite{piazza2014planar,li20202d, kaneti2021borophene,zhang2017two}. Different phases of bulk and two-dimensional (2D) boron allotropes, such as $\alpha$, $\beta$, and others, have been put forward in theoretical proposals \cite{Zhang2018Oblique,kong2021oblique, lopez2016electronic, gonzalez2008boron, piazza2014planar}. Although boron typically avoids forming chemical bonds to maintain a stable honeycomb lattice, it is feasible to generate a stable planar structure by integrating honeycomb with triangular units  \cite{tang2007novel, tang2009self}. This structure, known as $2B:Pmmn$, has two atoms per primitive unit cell and belongs to the orthorhombic crystal system with space group 59 ($Pmmn$). A novel Dirac material known as hydrogenated borophene, or borophane, was predicted \cite{xu2016hydrogenated} to exhibit Dirac properties with a remarkable Fermi velocity nearly twice that of graphene.

	Characterized by interesting properties different from those of graphene, borophene has emerged as a fascinating facet in the field of 2D materials. The importance of studying graphene and borophene lies in their different contributions to nanoscale physics and materials science, highlighting the need for further exploration of these materials \cite{jugovac2023coupling, liu2019borophene, hou2021borophene, wang2020activating, mortazavi2020machine}. By exploring the interfaces between 2D materials, one can discover a realm of intricate interactions and emergent phenomena \cite{das2015beyond, butler2013progress, ferrari2015science}. The key to unlocking new functionalities lies at these interfaces, where the unique properties of individual materials come together. The integration of graphene and borophene junctions has the potential to provide unprecedented electronic, structural, and transport properties \cite{xie2021chemistry, tan2017recent}. The study of this interface becomes not only a quest to understand the interaction of graphene and borophene, but also a gateway to exploit their combined potential for innovative applications in nanoscale devices and materials engineering \cite{ma2020review, usman2021bismuth, riazi2021ti}. Notable contributions from various studies highlight the importance of understanding this interface, with research on graphene  \cite{iqbal2023nanostructures, abdullah2017quantum, PhysRevB.108.245419, behura2019graphene} and borophene \cite{khalatbari2021spin, xu2023valley} junctions  serving as crucial steps.

Building on previous advances in understanding the transport properties of graphene (G) and borophene (B) in the presence of various potential barriers, we undertake a similar investigation focused on analyzing the behavior of two junctions, BGB and GBG. Specifically, we study the case where these two junctions encounter a single barrier and extend it to scenarios with multiple barriers, similar to a superlattice.
	Consequently, we have arrived at the key findings: In the case of a single barrier, our study has uncovered remarkable electronic transport properties. Indeed, regardless of the barrier width or height, perfect transmission is consistently observed at normal incidence in both junctions. However, the BGB junction has a {predominant} effect, showing the influence of the potential barrier, as reducing the barrier height results in a loss of transmission. We observe pronounced scattered transmission at normal incidence as long as the energy and transverse wave vector $k_y$ vary. In addition, we show that the single barrier induces resonances associated with the finite size of graphene and borophene. 
In the scenario involving multiple barriers (superlattice), we observe remarkable transport properties. Notably, the superlattice maintains perfect transparency for the BGB junction even at non-null incident angles, highlighting robust transmission. Resonance peaks, particularly notable in BGB, depict distinctive transport phenomena that are affected by the number and position of the cells. Specifically, as the energy varies, multiple transmission gaps emerge. The number, width, and position of these gaps can be modified by adjusting the number of cells, incident angle, and barrier characteristics.
	In the case of diffuse transport, we find {considerable variations in transmission probability, conductance and Fano factor}, highlighting the sensitivity of BGB and GBG junctions to physical parameters. In particular, we observed different behaviors between BGB and GBG junctions, especially in terms of how the conductance and Fano factor respond to variations in barrier height. In the case of ballistic transport, an intriguing relationship emerged between the minimum {scaled conductance} and the maximum Fano factor. This relationship demonstrated the controllability of these factors under certain conditions of the physical parameters. In addition, our analysis of the length ratio, also known as the geometric factor, revealed remarkable patterns. As the ratio approached infinity, the {scaled conductance} and the Fano factor converged to certain values, suggesting intriguing properties associated with this geometric configuration.

	\begin{figure}[t!]
		\centering
		\includegraphics[width=\linewidth]{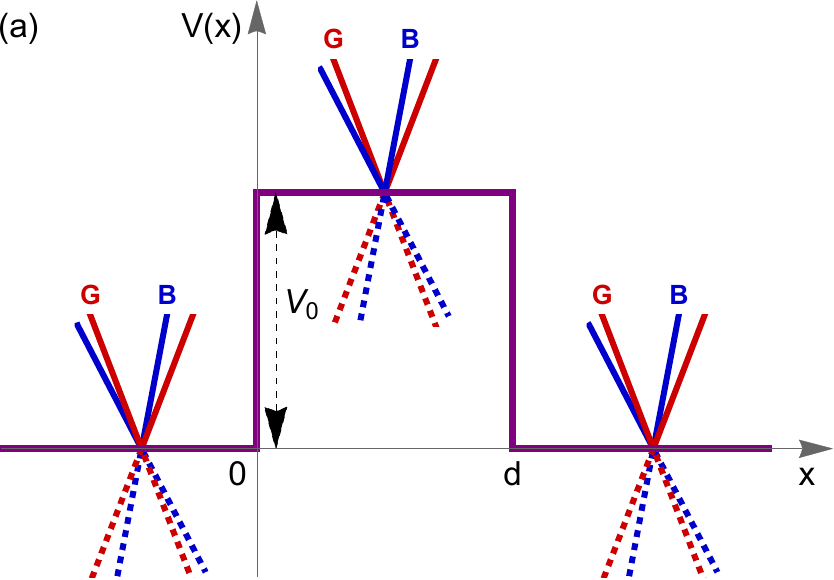}\\
		\includegraphics[width=\linewidth]{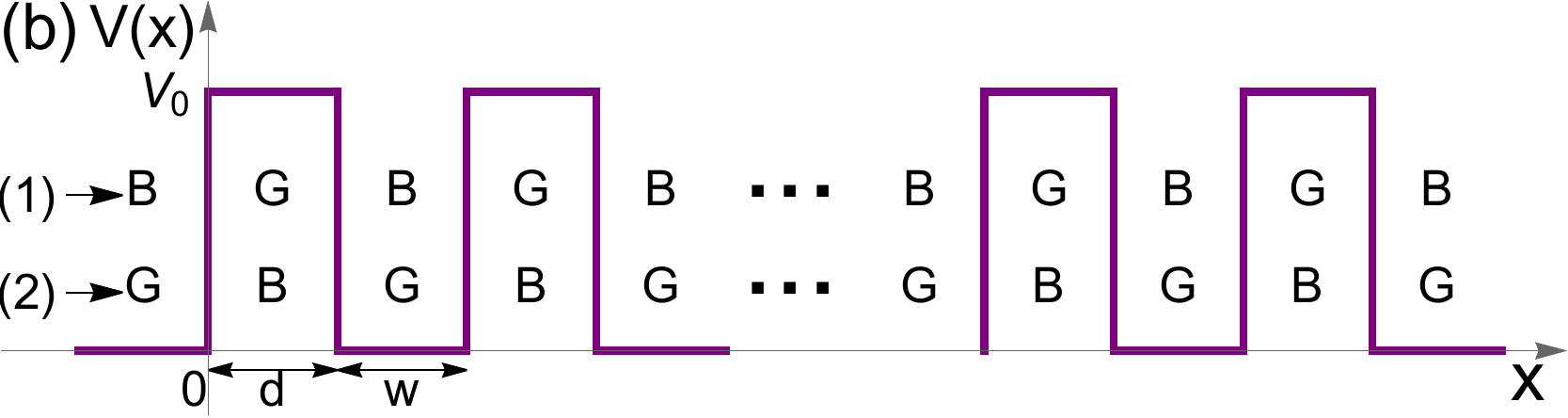}
		\caption{(Color online) {(a) Band dispersion relations around the Dirac point for BGB-junction and GBG-junction are plotted in each region of a rectangular potential barrier of height $V_0$ and width $d$. The blue tilted energy and red linear energy dispersions correspond to borophene (B) and graphene (G), respectively. (b) defines the 1D periodic potential of $(\text{BG})^n$ (case 1) and $(\text{GB})^n$ (case 2) superlattices, where $n$ is the number of cells, $d$ is the barrier width, and $w$ is the well width. We have set the period to $L=d+w$.}
		}\label{Schematicsystem}
	\end{figure}

	The paper is organized as follows. In Sec.~\ref{theoretical exploration}, we present a theoretical model describing fermions in borophene and graphene. In Sec.~\ref{TTT}, we use a mathematical formalism based on the continuity of spinors at interfaces and the transfer method to compute the transmission, conductance, and Fano factor. In Sec.~\ref{numerical results}, we numerically analyze our results by offering different discussions and comparisons with literature. We summarize our main conclusions in Sec.~\ref{conclusions}.
	
	%%%%%%%%%%%%%%%%%%%%%%%%%%%%%%%%%%%%%%%%%%%%%%%%%%%%%%%%%%%%%%%%%%%
	\section{Theoretical MODEL}\label{theoretical exploration}
	%%%%%%%%%%%%%%%%%%%%%%%%%%%%%%%%%%%%%%%%%%%%%%%%%%%%%%%%%%%%%%%%%%%

{In Fig.~\ref{Schematicsystem}, we illustrate the two configurations: borophene-graphene-borophene (BGB) and graphene-borophene-graphene (GBG). According to Fig. \ref{Schematicsystem}(a), one sees that both borophene and graphene exhibit linear dispersions typical of Dirac cones. These are visible within a small energy range known as the low-energy approximation \cite{feng2017dirac}.} 	
	To study the transport properties in both junctions and provide comparisons, we consider the periodic potential illustrated in Fig. \ref{Schematicsystem} (b) and given by 
	\begin{equation}
		V_j(x)=\left\{
		\begin{array}{ll} 
			V_0, & \ \text{if} \ (j-1)L\leq x \leq(j-1)L+d\\
			0, & \ \text{otherwise} 
		\end{array}
		\right.
		\label{E1}
	\end{equation}
	where $V_0$ is the barrier height, $d$ and $w$ are, respectively, the barrier and well widths, and $j$ is an integer, $1\leq j \leq n$, with $n$ rectangular barriers. The distance $L = d + w$ represents the width of the unit cell, which is composed of one barrier and one well.
	
	%%%%%%%%%%%%%%%%%%%%%%%%%%%%%%%%%%%%%%%%%%%%%
	\subsection{Borophene}
	%%%%%%%%%%%%%%%%%%%%%%%%%%%%%%%%%%%%%%%%%%%%%
	%%%%%%%%%%%%%%%%%%%%%%%%%%%%%%%%%%%%%%%%%%%%%
	%\subsubsection{Borophene}
	%%%%%%%%%%%%%%%%%%%%%%%%%%%%%%%%%%%%%%%%%%%%%
	We initiate our study with the low-energy continuum Hamiltonian for a monolayer  borophene in $j$th region, which describes a tilted anisotropic Dirac cone in 2D using three parameters \cite{sadhukhan2017anisotropic, nakhaee2018tight, Zhou2019Valley,zabolotskiy2016strain}
	\begin{equation}\label{Hamiltonian for borophene}
		H_{\text{B},j}=\hbar v_x k_x \sigma_x+\hbar v_y k_y \sigma_y+\hbar v_t k_y \sigma_0+V_{j}(x)\sigma_0,
	\end{equation}
	where $\sigma_x, \sigma_y$ are the Pauli matrices for the pseudospin representing the lattice degree of freedom, while $\sigma_0$ is the $2 \times$2 identity matrix, $(k_x,k_y)$ are the wave vector components, and the three distinct velocities are $\{v_x, v_y, v_t\}=\{0.86, 0.69, 0.32\}\times v_F$ \cite{sadhukhan2017anisotropic, nakhaee2018tight, Zhou2019Valley,zabolotskiy2016strain}, with $v_F=10^6 \text{m/s}$ is the Fermi velocity.
	We can easily show that tilted relation dispersion corresponds to Hamiltonian Eq. \eqref{Hamiltonian for borophene} reads as
	\begin{equation}
		E^s_{\text{B},j}(k)=V_j+\hbar v_t k_y+s \hbar \sqrt{v_x^2 k_{x,j}^2+v_y^2 k_y^2},
	\end{equation}
	in which $s=\pm$ indicate the conduction and valence bands, respectively. This relation is presented in  Fig.~\ref{Schematicsystem} with blue lines.

	By taking into account the conservation of the transverse wave vector $k_y$ and using the eigenvalue equation $H_{B,j}\Psi_{B,j} = E_{B,j}\Psi_{B,j}$ for a given region $j$, we show that the associated eigenspinors can be written in matrix form as
	\begin{equation}
		\Psi_{\text{B},j}(x, y)=M_{\text{B},j}(x)\cdot
		\binom{a_{B}}{b_{B}}  e^{ik_y y},
	\end{equation}
	where $M_{\text{B},j}(x)$ is given by
	\begin{equation}
		M_{\text{B},j}(x)=\begin{pmatrix}
			e^{i k_{x,j} x} & e^{-i k_{x,j} x} \\
			f_j^{+}e^{i k_{x,j} x} & f_j^{-}e^{-i k_{x,j} x}
		\end{pmatrix},
	\end{equation}
	and the wave vector along $x$-direction reads as
	\begin{equation}
		k_{x,j}=\frac{1}{\hbar v_x}\sqrt{\varepsilon_{\text{B},j}^2-2\hbar v_t\varepsilon_{\text{B},j}k_y+\hbar^2v_t^2k_y^2-\hbar^2v_y^2k_y^2}.
	\end{equation}
	Here, we have set  the quantities $f_j^{\pm}=\frac{i\hbar v_y k_y\pm\hbar v_x k_{x,j}}{\varepsilon_{\text{B},j}-\hbar v_t k_y}$, and $E_{\text{B},j}-V_j=\varepsilon_{\text{B},j}$.
	
	%%%%%%%%%%%%%%%%%%%%%%%%%%%%%%%%%%%%%%%%%%%%%
	\subsection{Graphene}
	%%%%%%%%%%%%%%%%%%%%%%%%%%%%%%%%%%%%%%%%%%%%%
	
	For graphene regions, an electron in the presence of an electrostatic potential is described by the following Hamiltonian \cite{katsnelson2006chiral}
	\begin{equation}\label{Hamiltonian for graphene}
		H_{\text{G},j}=v_F \vec\sigma\cdot\vec \pi+V_{j}(x)\sigma_0,
	\end{equation}
	where $\pi=p_{x}+ip_{y}$ is the in-plan momentum and $\vec\sigma=\{\sigma_x,\sigma_y\}$ are the Pauli matrices.
	The corresponding linear band dispersion
	near the Dirac points takes the form
	\begin{equation}
		E^s_{\text{G},j}(k)=V_j+s\hbar v_F \sqrt{q_{x,j}^2+k_y^2},
	\end{equation}
	where $q_{x,j}$ and $k_y$ are wave vector components. Here,
	$s=\pm$ denotes the conduction and the valence band in graphene dispersion, which is illustrated in Fig.~\ref{Schematicsystem} (red lines).
	
	Regarding the eigenspinors, from the eigenvalue equation $H_{G,j}\Psi_{G,j} = E_{G,j}\Psi_{G,j}$, we show that they can be cast as follows
	\begin{equation}
		\Psi_{\text{G},j}(x, y)=M_{\text{G},j}(x)\cdot
		\binom{a_G}{b_G}
		 e^{ik_y y},
	\end{equation}
	and we have
	\begin{equation}
		M_{\text{G},j}(x)=\begin{pmatrix}
			e^{iq_{x,j} x} & e^{-iq_{x,j} x} \\
			g_{j}  e^{iq_{x,j} x} & g_{j}^{*} e^{-iq_{x,j} x}
		\end{pmatrix}
	\end{equation}
	with the complex number $g_j=\frac{1}{g_j^{*}}=\text{sgn}(\varepsilon_{\text{G},j})\frac{q_{x,j}\pm ik_y}{\varepsilon_{\text{G},j}}$, $E_{\text{G},j}-V_j=\varepsilon_{\text{G},j}$, and the wave vector $q_{x,j}$ is defined by
	\begin{equation}
		q_{x,j}=\sqrt{\left(\frac{\varepsilon_{\text{G},j}}{\hbar v_F}\right)^{2}-k_y^2}.
	\end{equation}
	%It is important to highlight
	Note  that the wave vectors $k_{x,j}$ and $q_{x,j}$ are both depend on the incident angle $\theta_j$ within the $j$th potential barrier. This dependence is expressed through the relations $\theta_j=\arcsin\frac{k_y}{k_{x,j}}$ for BGB and $\theta_j=\arcsin\frac{k_y}{q_{x,j}}$ for GBG.
	
	\vspace{2mm}
	%%%%%%%%%%%%%%%%%%%%%%%%%%%%%%%%%%%%%%%%%%%%%%%%%
	\section{TRANSPORT quantities}\label{TTT}
	%%%%%%%%%%%%%%%%%%%%%%%%%%%%%%%%%%%%%%%%%%%%%%%%
	By ensuring the continuity of the wave function across the various boundaries between barriers and wells, we can establish the transfer matrix for the periodic structures BGB and GBG \cite{wang2010electronic, wang2014transfer, li2009generalized}, each composed of $n$ unit cells, respectively,
	\begin{align}
		&\label{transfermatrixBGB}
		M_{n,\text{BGB}}=M_{\text{B},1}(0)^{-1}\Omega_{\text{BGB}}^n M_{\text{B},1}(nL),
		\\
		&\label{transfermatrixGBG}
		M_{n,\text{GBG}}=M_{\text{G},1}(0)^{-1}\Omega_{\text{GBG}}^n M_{\text{G},1}(nL),
	\end{align}
	where the matrix $\Omega_{\text{BGB/GBG}}^n$ are given by
	%\begin{subequations}
	\begin{align}\label{omegaBGBgbg}
		\Omega_{\text{BGB}}&=M_{\text{G},2}(L)M_{\text{G},2}(d)^{-1},\\
		\Omega_{\text{GBG}}&=M_{\text{B},2}(L)M_{\text{B},2}(d)^{-1}. \label{om22}
	\end{align}
	To go further, we establish the dispersion relations corresponding to BGB and GBG junctions. Utilizing Bloch's theorem for a period $L$ and the dispersion relation $\cos(K_{x}L)=\frac{1}{2}\tr(M_{n,\text{BGB/GBG}})$ for $n$ cells, we arrive at the following expressions for BGB and GBG, respectively:
	\begin{widetext}
		\begin{align}
			\cos(K_xL)&=\alpha\left[\left(g_2-f_1^{-}\right)\left(1+f_1^{+}g_2\right)\cos(d k_{x,1}+w q_{x,2})+\left(f_1^{+}-g_2\right)\left(1+f_1^{-}g_2\right)\cos(d k_{x,1}-w q_{x,2})\right]\label{EqBSBGB},\\
			\cos(K_xL)&=\beta\left[\left(g_1-f_2^{-}\right)\left(1+f_2^{+}g_1\right)\cos(d q_{x,1}+w k_{x,2})+\left(f_2^{+}-g_1\right)\left(1+f_2^{-}g_1\right)\cos(d q_{x,1}-w k_{x,2})\right] \label{EqBSGBG},
		\end{align}
	\end{widetext}
	where we have set the parameters $\alpha=\frac{1}{\left(1+g_2^2\right)\left(f_1^{+}-f_1^{-}\right)}$ and  $\beta=\left(1+g_1^2\right)\left(f_2^{+}-f_2^{-}\right)$,  $K_x$ is the Bloch wave vector.
	
\begin{figure}[t!]
	\vspace{0.cm}
	\centering\graphicspath{{./Figures/}}
	\includegraphics[width=\linewidth]{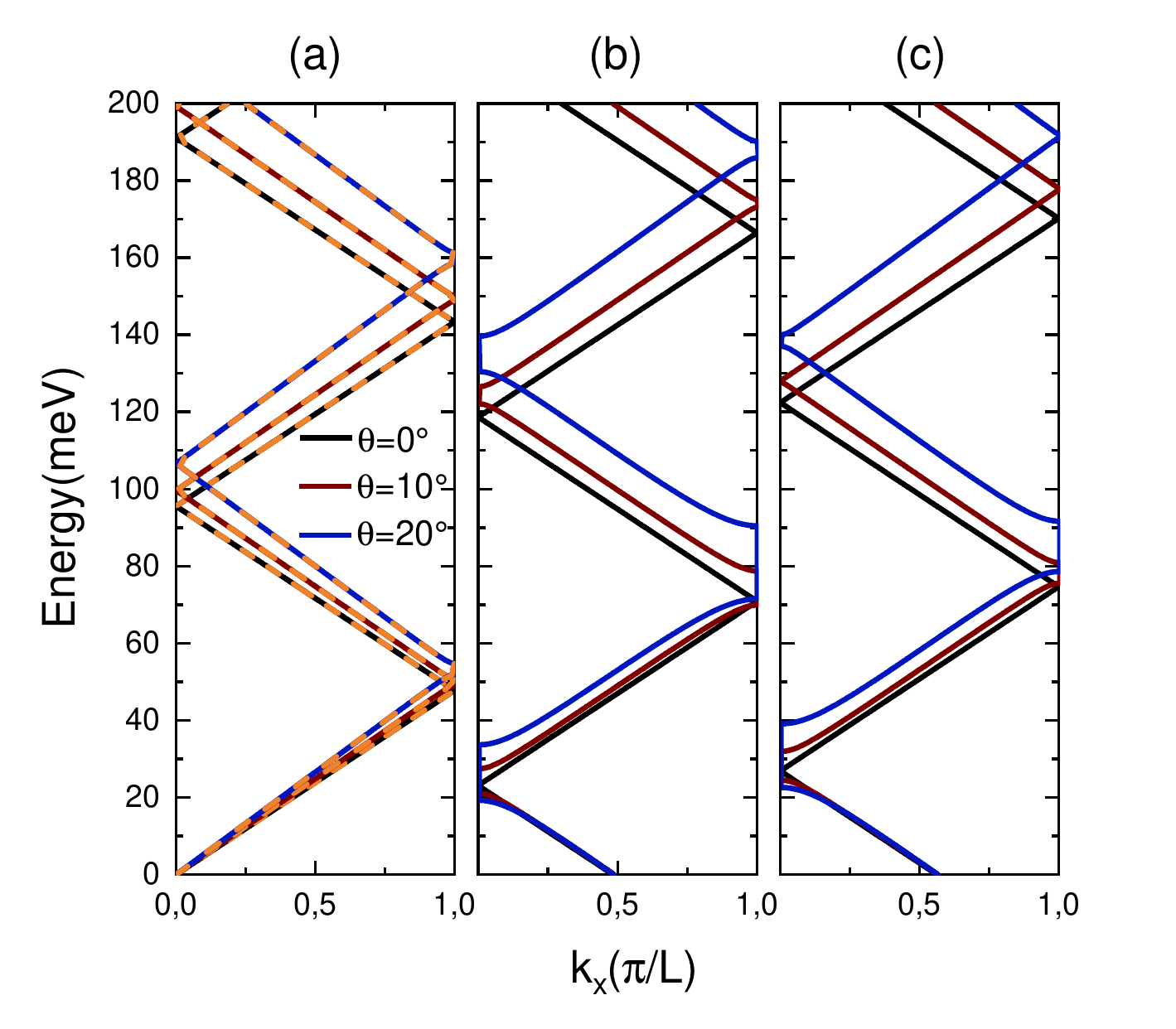}
	\caption{(Color online) (a) The band structures of BGB (solid lines) and GBG (dashed orange lines) junctions for $V_0=0$ and $d=w=20$ nm, but BGB (b) and GBG (c) for $V_0=50$ meV.}\label{Bandstructure}
\end{figure}

	In Fig. \ref{Bandstructure}, we illustrate the band structures for BGB (Eq. \eqref{EqBSBGB}) and GBG (Eq. \eqref{EqBSGBG}) under appropriate conditions of the physical parameters. This sheds light on the impact of the incident angle on the mini-zone boundary. In particular, enlarged depictions of the band structures near the wave vector $\pi/L$ are presented for BGB and GBG junctions by choosing the barrier height between the two layers $V_0=0$ (a) and $V_0=50$ meV (b, c), where $d=w=20$ nm. In Fig. \ref{Bandstructure}(a) for $V_0=0$, a notable observation arises in the band structures of BGB and GBG junctions. Regardless of the values taken by the incident angle $\theta$, both BGB and GBG band structures remain unchanged with no gap opening, unlike the findings observed in graphene superlattices \cite{XU2015188superlattices}. {This observation underscores the critical role of barrier height in scenarios involving non-zero potential barriers, which will be discussed in the following sections.}
	  Subsequently, in Figs. \ref{Bandstructure}(b, c), we observe that the energy separation around $\pm \pi/L$ depends on both the wave vector and incident angle. Notably, at $\theta=0^\circ$, there is no gap opening at the center of the mini-zone boundary for both the two junctions. However, at $\theta=10^\circ$ and $20^\circ$, the original Dirac cones have upward shifts, and the gap exhibits sensitivity to the incident angle.

In order to determine the transmission coefficient for $n$ barriers, we introduce the Chebyshev polynomials  \cite{mason2002chebyshev, covaci2010efficient}. Specifically, we map the transfer matrices  Eqs.  \eqref{transfermatrixBGB} and \eqref{transfermatrixGBG} as
\begin{widetext}\label{}
	\begin{align}
		M_{n,\text{BGB}}&=M_{\text{B},1}(0)^{-1}\begin{pmatrix}
			U_{n-1}(\omega_{\text{BGB}})\Omega_{\text{BGB},11}-U_{n-2}(\omega_{\text{BGB}}) & U_{n-1}(\omega_{\text{BGB}})\Omega_{\text{BGB},12} \\
			U_{n-1}(\omega_{\text{BGB}})\Omega_{\text{BGB},21}& U_{n-1}(\omega_{\text{BGB}})\Omega_{\text{BGB},22}-U_{n-2}(\omega_{\text{BGB}})
		\end{pmatrix} M_{\text{B},1}(nL),\\
		&=\begin{pmatrix}
			M_{n,\text{BGB},11} & M_{n,\text{BGB},12}\\
			M_{n,\text{BGB},21} & M_{n,\text{BGB},22}
		\end{pmatrix},\\
		M_{n,\text{GBG}}&=M_{\text{G},1}(0)^{-1}\begin{pmatrix}
			U_{n-1}(\omega_{\text{GBG}})\Omega_{\text{GBG},11}-U_{n-2}(\omega_{\text{GBG}}) & U_{n-1}\Omega_{\text{GBG},12}(\omega_{\text{GBG}}) \\
			U_{n-1}(\omega_{\text{GBG}})\Omega_{\text{GBG},21}& U_{n-1}(\omega_{\text{GBG}})\Omega_{\text{GBG},22}-U_{n-2}(\omega_{\text{GBG}})
		\end{pmatrix}  M_{\text{G},1}(nL),\\
		&=\begin{pmatrix}
			M_{n,\text{GBG},11} & M_{n,\text{GBG},12}\\
			M_{n,\text{GBG},21} & M_{n,\text{GBG},22}
		\end{pmatrix},
	\end{align}
\end{widetext}
where $U_n(x)= \frac{\sin[(n+1)x]}{\sin x}$ represent Chebychev polynomials of the second kind, $\omega_{\text{BGB/GBG}}=\arccos\left[\frac{1}{2}\tr\left(M_{1,\text{BGB/GBG}}\right)\right]$, and $\Omega_{\text{BGB/GBG},ij}$ are the $(i,j)$ elements of the single period matrices Eqs. (\ref{omegaBGBgbg}, \ref{om22}). This process results in acquiring the transmission coefficient associated with the $n$ barrier, as
\begin{align}
	&T_{\text{BGB},n}=\frac{1}{\left|M_{n,\text{BGB},11}\right|^{2}},\label{T1}\\ &T_{\text{GBG},n}=\frac{1}{\left|M_{n,\text{GBG},11}\right|^{2}}.\label{T2}
\end{align}

Based on the Landauer-B\"uttiker  formula \cite{buttiker1985generalized} and the transmission probabilities Eqs. (\ref{T1}, \ref{T2}), we compute the conductance as a function of the incident angle $\theta$. This is 
\begin{equation}
	G_{\text{BGB/GBG},n}=G_0 \int_{-\pi / 2}^{\pi / 2} T_{\text{BGB/GBG},n} \cos \theta d \theta,
\end{equation}
where $\theta=\arccos\frac{Q_x}{E_F}$, with $Q_x=k_x$ for borophene and $Q_x=q_x$ for graphene. Here, $G_0=2 e^2 E_F D /(\pi \hbar)$ is the conductance unit, and $D$ is the sample size in the $y$-direction. Finally, the Fano factor is given by
 \cite{Beenakker2006SubPoissonian, lima2018tuning}
\begin{equation}
	F_{\text{BGB/GBG},n}=\frac{\int_{-\pi / 2}^{\pi / 2} T_{\text{BGB/GBG},n}(1-T_{\text{BGB/GBG},n}) \cos \theta d \theta}{\int_{-\pi / 2}^{\pi / 2} T_{\text{BGB/GBG},n} \cos \theta d \theta}.
\end{equation}

%%%%%%%%%%%%%%%%%%%%%%%%%%%%%%%%%%%%%%%%%%%%%%%%%%
\section{RESULTS AND DISCUSSIONs}
\label{numerical results}
%%%%%%%%%%%%%%%%%%%%%%%%%%%%%%%%%%%%%%%%%%%%%%%%%
Having established the analytical findings, we shift to conducting numerical analysis on the transport properties linked to the GBG and BGB junctions across different scenarios. Our goal is to provide a comprehensive understanding of the electronic transport phenomena that occur at these two junctions. To accomplish this, we consider single and multiple barriers as distinct entities.

%%%%%%%%%%%%%%%%%%%%%%%%%%%%%%%%%%%%%%%%%%%%%%%%%
\subsection{Single barrier (one cell)}
\label{One barrier}
\begin{figure}[t!]
	\centering
	\includegraphics[width=\linewidth]{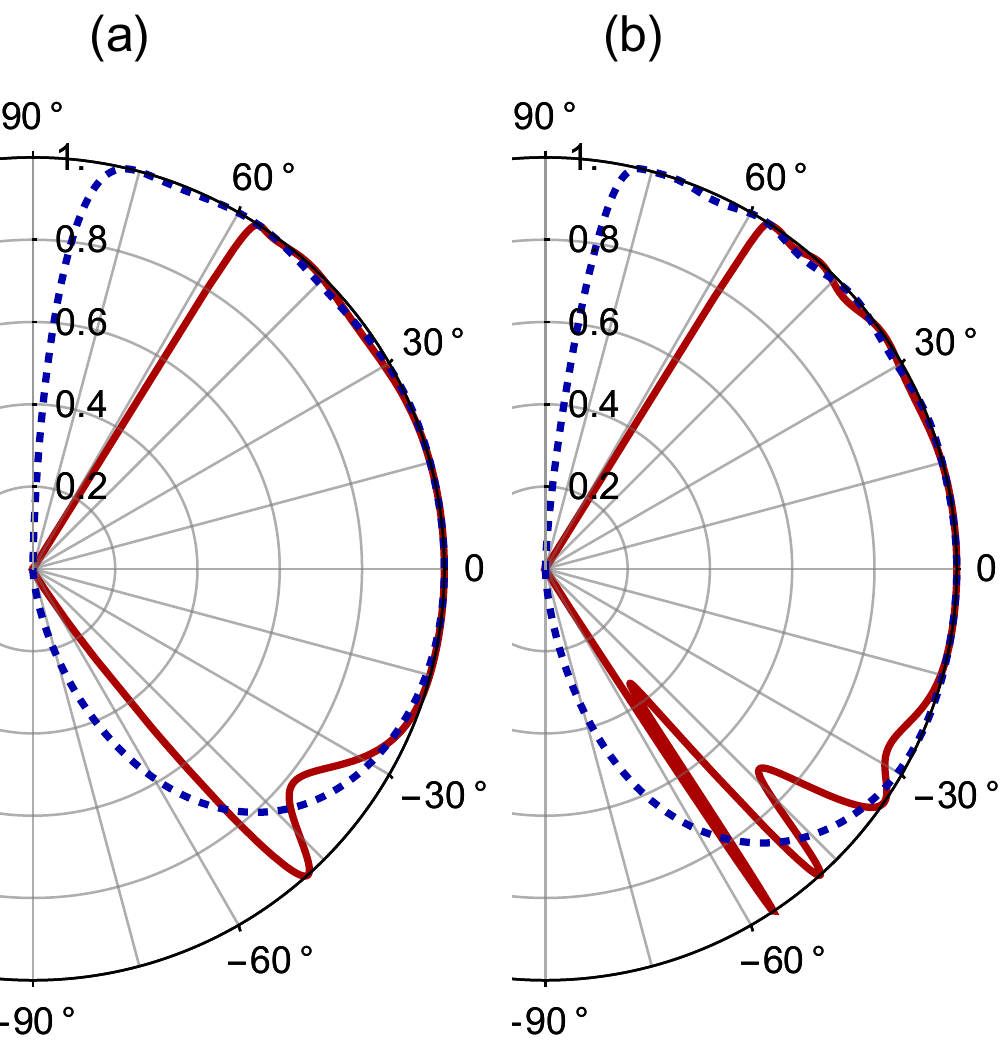}
	\caption{{(Color online) Angular dependence of transmission probability for BGB and GBG junctions with $E=82.49$ meV, $V_0=0$, and two barrier widths: (a) $d=50$ nm, (b) $d=100$ nm. Red curves represent BGB, and blue curves represent GBG.}}  \label{Tpolar0}
\end{figure}

%%%%%%%%%%%%%%%%%%%%%%%%%%%%%%%%%%%%%%%%%%%%%%%%%
In Fig.~\ref{Tpolar0}, we present the transmission probabilities as a function of the incident angle $\theta$ in the absence of potential  barrier  ($V_0 = 0$) for two barrier widths: (a) $d=50$ nm and (b) $d=100$ nm.  Remarkably, perfect transmission is observed under normal incidence for both BGB and GBG junctions, regardless of the particular value of $d$,  resulting in signature of Klein tunneling. i.e., $T=1$. This robust behavior highlights the exceptional transport properties of these structures, especially concerning the normal incident angle $(\theta=0)$ and $V_0 = 0$. This effect remains consistent even when the potential barrier is non-zero. In contrast to the findings in \cite{katsnelson2006chiral}, the transmissions exhibit asymmetry for both junctions at non-normal incidence. Furthermore, perfect transmission occurs over {a wide range} of incident angles associated with each junction.
This phenomenon of perfect tunneling can be explained by invoking the principle of pseudospin conservation between the energy states of graphene and borophene.

\begin{figure}[t!]
	\centering
	\includegraphics[width=\linewidth]{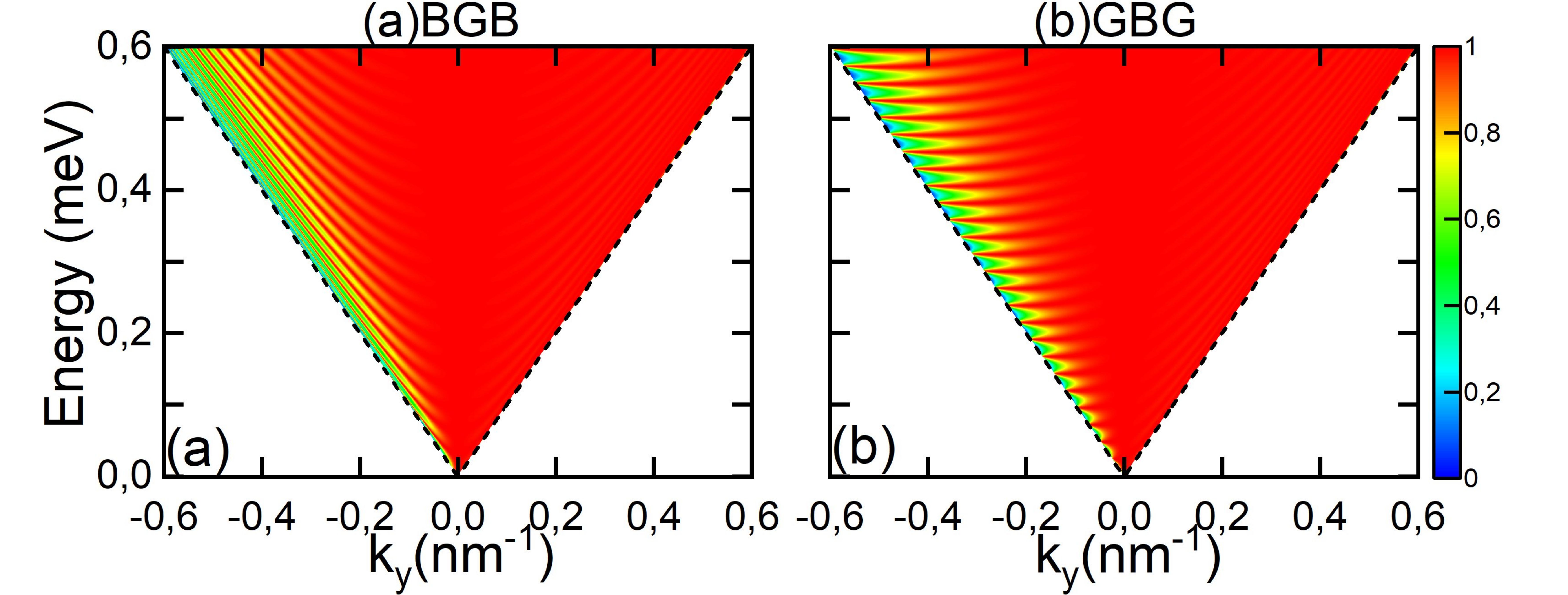}
	\caption{{(Color online) Contour plot of the transmission probability as a function of the energy and transverse wave vector for $V_0=0$ and  $d=100$ nm with BGB (a) and GBG (b). The dashed black lines represent the band energy for graphene.}}\label{Teky0BGBGBG}
\end{figure}

Fig.~\ref{Teky0BGBGBG} shows the contour plot of  the transmission probabilities versus the energy $E$ and transverse wave vector $k_y$ for $V_0 = 0$ and $d = 100$ nm. It depicts pronounced scattered transmission under normal incidence, a phenomenon strongly influenced by the dimensions of both graphene and borophene. This observation underscores the dynamic interplay between the structural dimensions of graphene and borophene. Notably, a consistent finding in both GBG and BGB configurations is that the transport primarily occurs through graphene. A similar behavior has been previously studied in \cite{abdullah2017quantum} by focusing pristine AA bilayer graphene sandwiched between two regions of a single layer of graphene.

\begin{figure}[tbh]
	\centering
	\includegraphics[width=\linewidth]{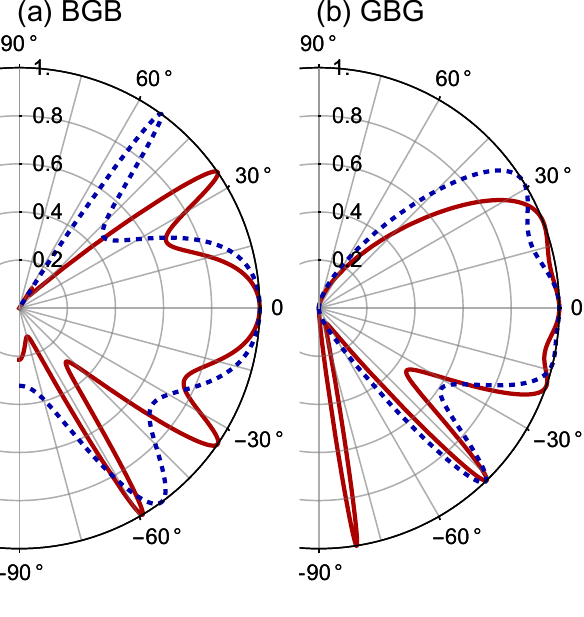}
	\caption{(Color online) The same as in Fig.~\ref{Tpolar0}(b) but now for $V_0=200$ meV (red curves) and  $V_0=285$ meV (blue curves).}\label{Tpolar}
\end{figure}
\begin{figure}[tbh]
	\centering
	\includegraphics[width=\linewidth]{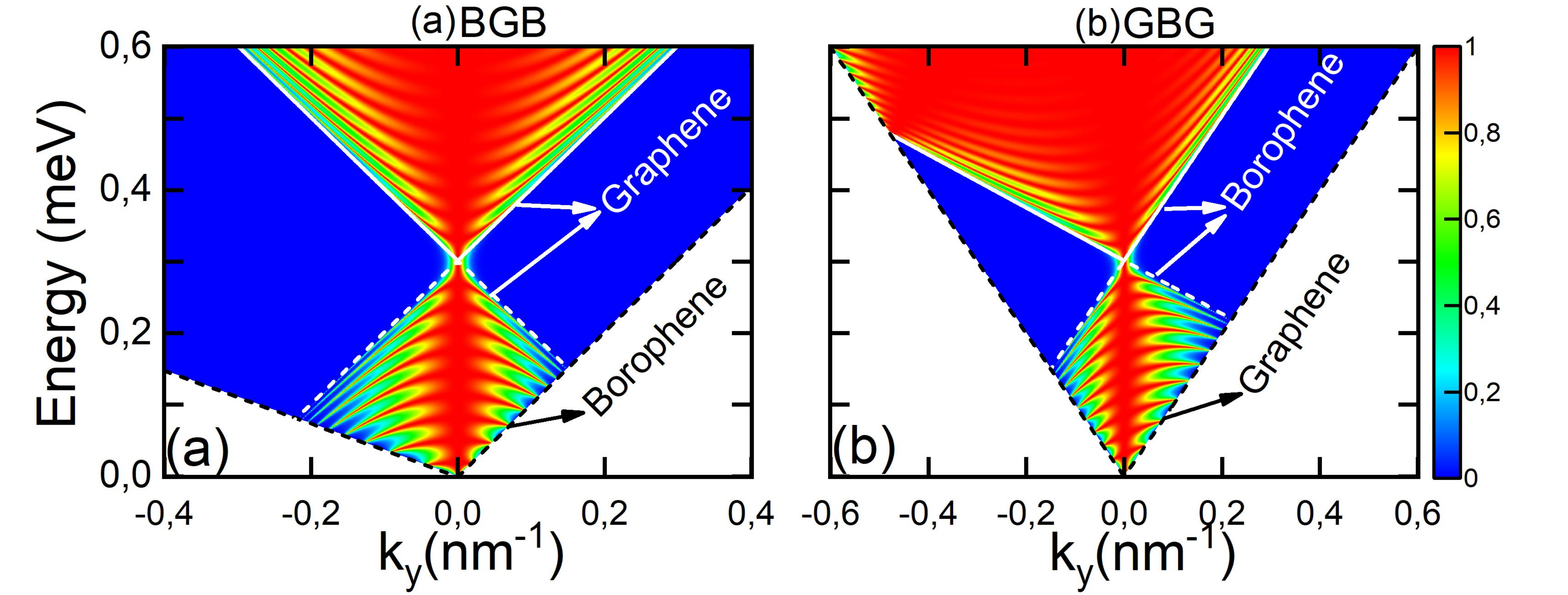}
	\caption{(Color online) The same as in Fig.~\ref{Teky0BGBGBG} but now for $V_0=0.3$ eV. The white and black lines represent the band inside and outside the barrier.}\label{TekyBGBGBG}
\end{figure}

While the previous analysis focused on junction behaviors independent of potential barriers, our investigation now extends to evaluating the consequences under the influence of various potential barriers.
In Fig.~\ref{Tpolar}, we show the angular dependence of transmission probability (a) BGB and (b) GBG, but now in the presence of the potential barrier ($V_0\not=0$). The discovery shows distinct patterns in electronic transport when Dirac fermions are normally incident at these junctions.
Particularly, we observe a perfect transmission at normal incidence, indicating distinct characteristics of Dirac fermions within these heterostructures. For the BGB junction with $V_0=285$ meV, the results align with those obtained in \cite{katsnelson2006chiral} by studying monolayer graphene. However, {a large shift} occurs when the barrier height is reduced to $200$ meV, resulting in vanishing transmission. 
The asymmetry in transmission at non-normal incidence for the two junctions is consistently achieved, mirroring the previous scenario where \( V_0 = 0 \). 
Furthermore, these resonances emerge even when the incident energy is below the barrier height, a phenomenon grounded in the conservation of pseudospin, directly associated with the Klein paradox in quantum electrodynamics \cite{katsnelson2006chiral}. The positions and numbers of these resonances primarily hinge on the barrier height and the specific junction.

	A contour plot of transmission probabilities as a function of energy and transverse wave vector is shown in Fig.~\ref{TekyBGBGBG} for $V_0=0.3$ eV and $d=100$ nm, such that BGB (a) and GBG (b). This figure shows a $(k_y, E)$ plane, identifying distinct quantifiable regions that signify different modes both inside and outside the junctions. {Remarkably, there is an asymmetry in the transmissions at $k_y=0$, which is 
	 	 due to the absence of valley equivalence \cite{van2013four}.}
	  In the presence of the potential barrier in the central region induces resonances in the transmission probabilities for energies in the range $E < V_0$. This phenomenon is associated with the finite size of graphene and borophene, as well as the presence of charge carriers with different chirality. Importantly, there is a noticeable difference between GBG and BGB.
	 This can be attributed to the lack of valley equivalence, where carriers scattering from borophene to graphene differ from those scattering in the opposite direction. 
	It is noteworthy that at $k_y=0$, Klein tunneling is observed in graphene \cite{katsnelson2006chiral} through a potential barrier, even at low applied potentials of $E<V_0$. Also, it emerges here for both BGB and GBG junctions, showing that the barrier consistently remains perfectly transparent.

	%%%%%%%%%%%%%%%%%%%%%%%%%%%%%%%%%%%%%%%%%%%%%%%%%%%%%%%
	\subsection{Multiple barrier ($n$ cells)}
	%%%%%%%%%%%%%%%%%%%%%%%%%%%%%%%%%%%%%%%%%%%%%%%%%%%%%%%
	\begin{figure}[t!]
		\vspace{0.cm}
		\centering\graphicspath{{./Figures/}}
		\includegraphics[width=\linewidth]{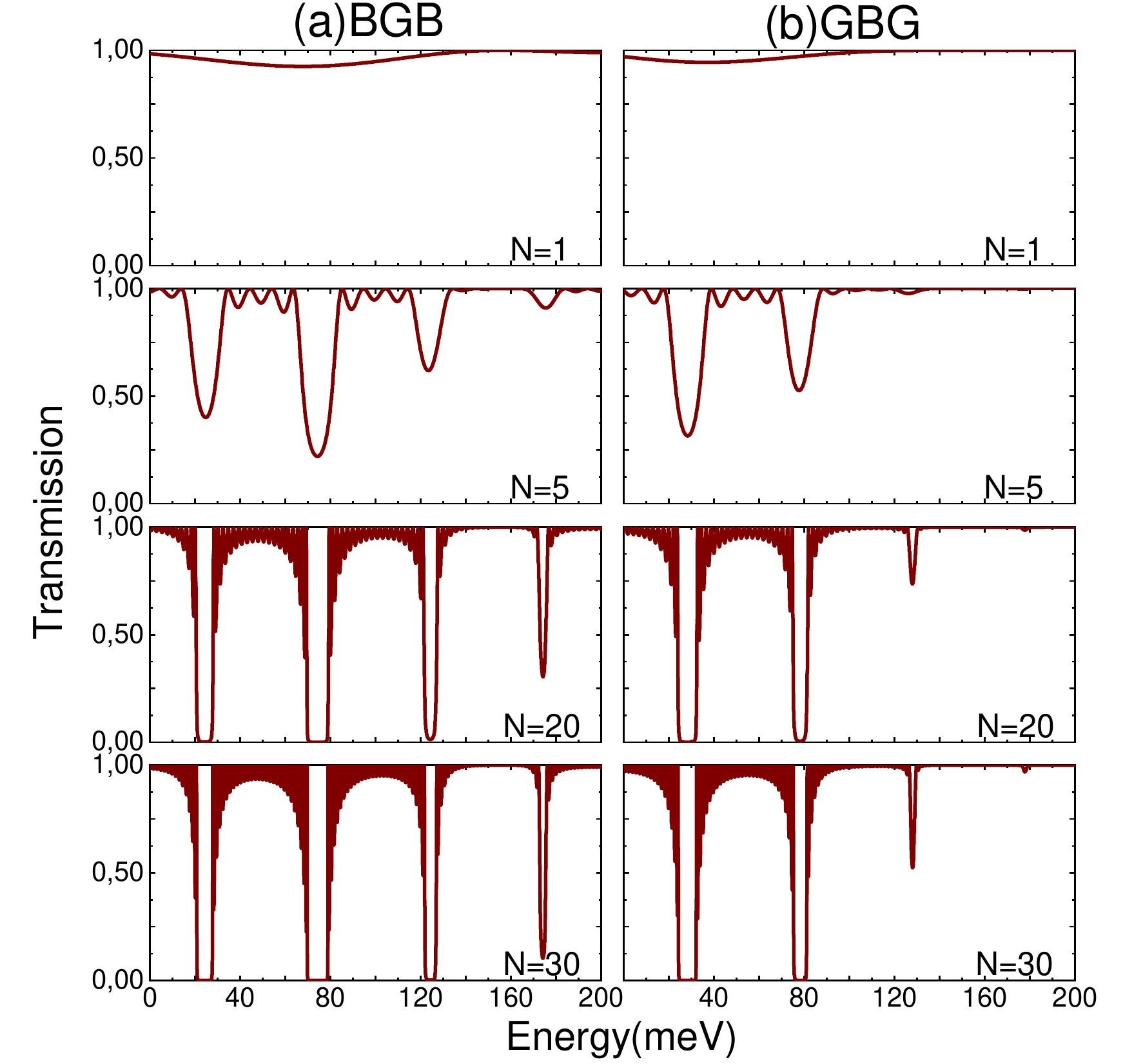}\
		\caption{(Color online) Transmission probability as a function the energy for BGB (a) and GBG (b), with $V_0=50$ meV,  $d=w=20$ nm,  $\theta_0=10^\circ$, and different numbers of cells $n$.}\label{TransmissioninE}
	\end{figure}
	We recall that Fig.~\ref{Schematicsystem} (c) presents the 1D periodic potential barrier, denoted by $(AB)^n$, with $n$ being the number of cells. It provides a visual representation of the repeating potential barriers inherent in these superlattices. Also, it serves as a key to exploring the complex interplay between periodic potential structures and electron transport phenomena within the studied superlattices. The periodicity of potential barriers, evident in the representation, is a crucial aspect influencing electron behavior in these systems.

	To show how the number of cells $n$ affects the transport properties, we plot transmission as a function of energy for different values of $n$ in Fig. \ref{TransmissioninE}. For $n=1$, we observe that the transmission is almost perfect, with a slight difference between the two junctions.  The transmission shows a consistent pattern as $n$ increases, suggesting resilience to fluctuations in $n$. 
	{Our results are in good agreement with those developed in other studies
	on  graphene subjected to multiple electrostatic barriers \cite{PhysRevLett.132.056204,dakhlaoui2021quantum, dakhlaoui2021modulating}.}
	We see that, despite increasing $n=5, 20,30$, the positions and numbers of resonant peaks indicate that the BGB junction retains complete transparency even at an incidence angle of $\theta=10^\circ$. Crucially, this transparency remains unaffected by variations in well widths. Resonances of the same kind are found for both junctions, with resonances in BGB being more {prominent} than in GBG.
	To compare with our results, it is worth noting that at normal incidence, a structure resembling monolayer graphene consistently remains perfectly transparent \cite{bai2007klein}. At $n=30$, we observe that {transmission gaps} begin to emerge. Indeed, the BGB junction exhibits three gaps, whereas the GBG junction has two. All these gaps fall within the same energy range, as depicted in the band structure shown in 
	Fig.~\ref{Bandstructure}.

\begin{figure}[t!]
		\centering
		\includegraphics[width=\linewidth]{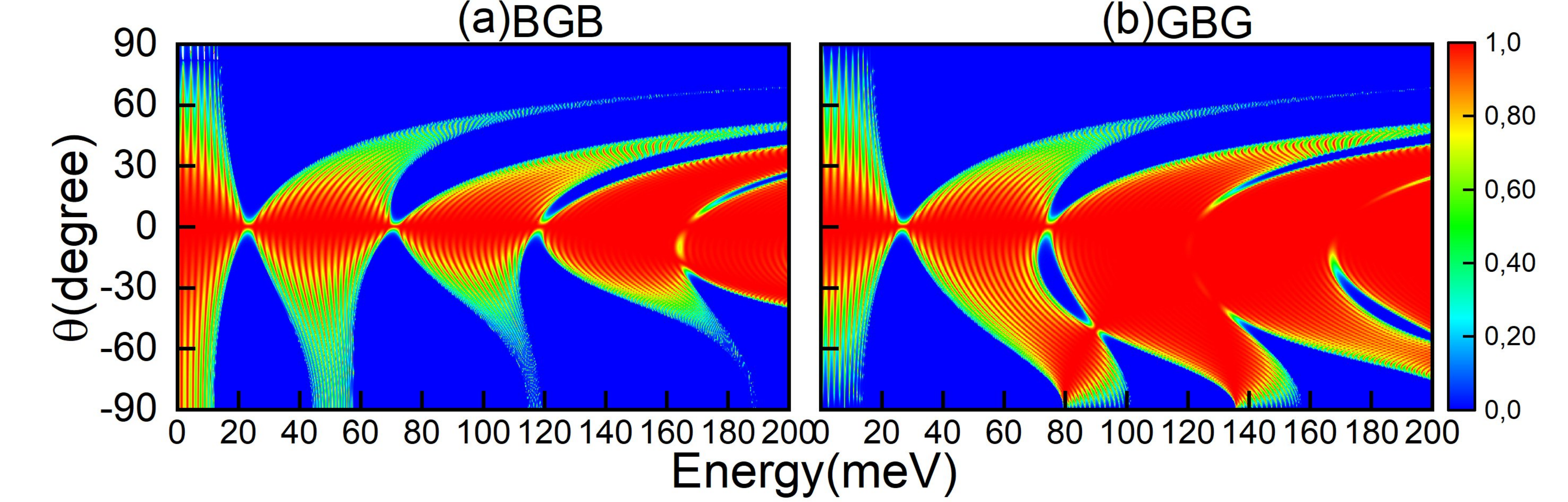}
		\caption{(Color online) Contour plot of the transmission probability as a function of the energy and incident angle for  BGB (a) and GBG  (b) through $N=20$ barriers, with $V_0=50$ meV and $d=w=20$ nm.}\label{transEPhi}
	\end{figure}

	In Fig.~\ref{transEPhi}, we plot transmission as a function of the energy and incident angle   in BGB (a) and GBG (b) junctions through $n=20$ barriers, with $V_0=50$ meV and $d=w=20$ nm. We find that the properties of graphene, which are widely known, differ {drastically} from those of borophene-graphene superlattices. Unlike graphene, where the directions of the wave vector and the group velocity are collinear, our results in BGB and GBG show unique patterns in how electrons cross barriers, depending on both energy $E$ and angle $\theta$. Anisotropic dispersion in borophene superlattices introduces noncollinear directions and gives access to valley birefringence \cite{xu2023valley}, whereas Klein tunneling in graphene is valley-degenerate and occurs mainly at normal incidence \cite{XU2015188superlattices}.
	Our results indicate that, in contrast to graphene, where certain energies lead to a complete blocking of electron motion, in the borophene arrangement, certain energies allow electrons to pass through barriers more freely. This effect is particularly pronounced in the GBG junction for energies above $80$ meV, while the behavior of the BGB junction is different.
	Note that for the BGB junction, the appearance of resonances in the transmission at $E=80$ meV and $E=136$ meV occurs only when the incident angle is negative. This is in contrast to the results obtained in \cite{xu2023valley, XU2015188superlattices}, where the resonances appear only for $E<20$ meV. Under the same condition and at $E=50$ meV, the transmission differs from zero for the BGB junction.

	%%%%%%%%%%%%%%
	%%%%%%%%%%%%%%
	%%%%%%%%%%%%%%
	\begin{figure}[t!]
		\centering
		\includegraphics[width=\linewidth]{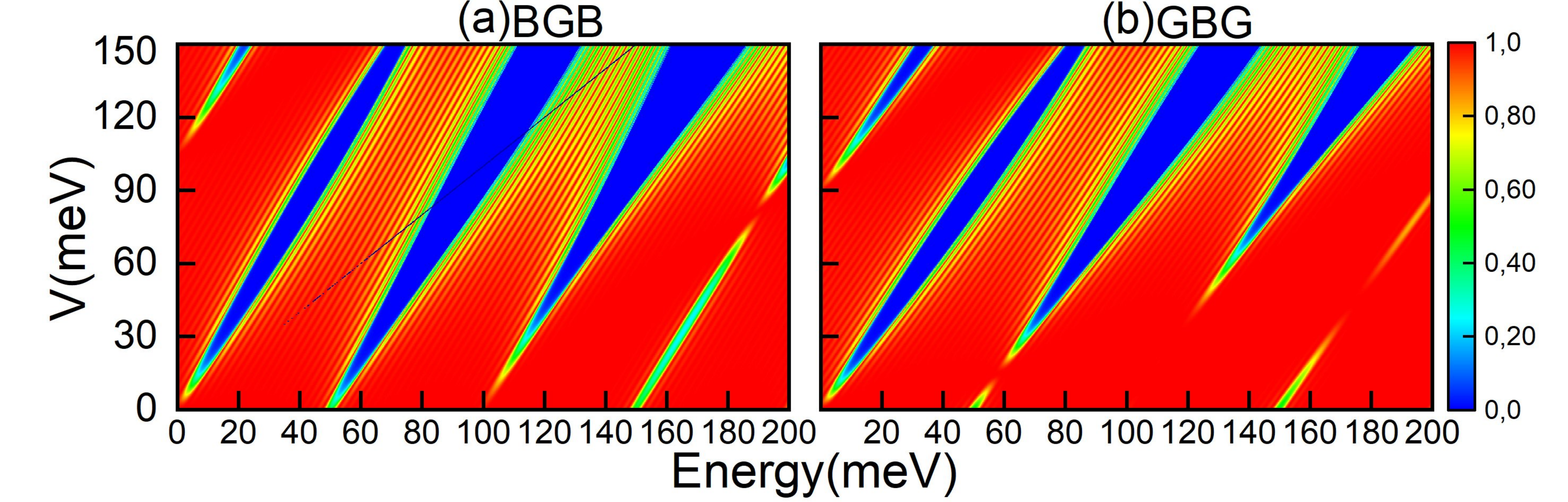}
		\caption{(Color online) The same as in Fig.~\ref{transEPhi} but now as a function of the energy and the barrier height $V_0$ for $\theta=10^\circ$ .}\label{transEV}
	\end{figure}
	%%%%%%%%%%%%%%
	%%%%%%%%%%%%%%
	\begin{figure}[t!]
		\centering
		\includegraphics[width=\linewidth]{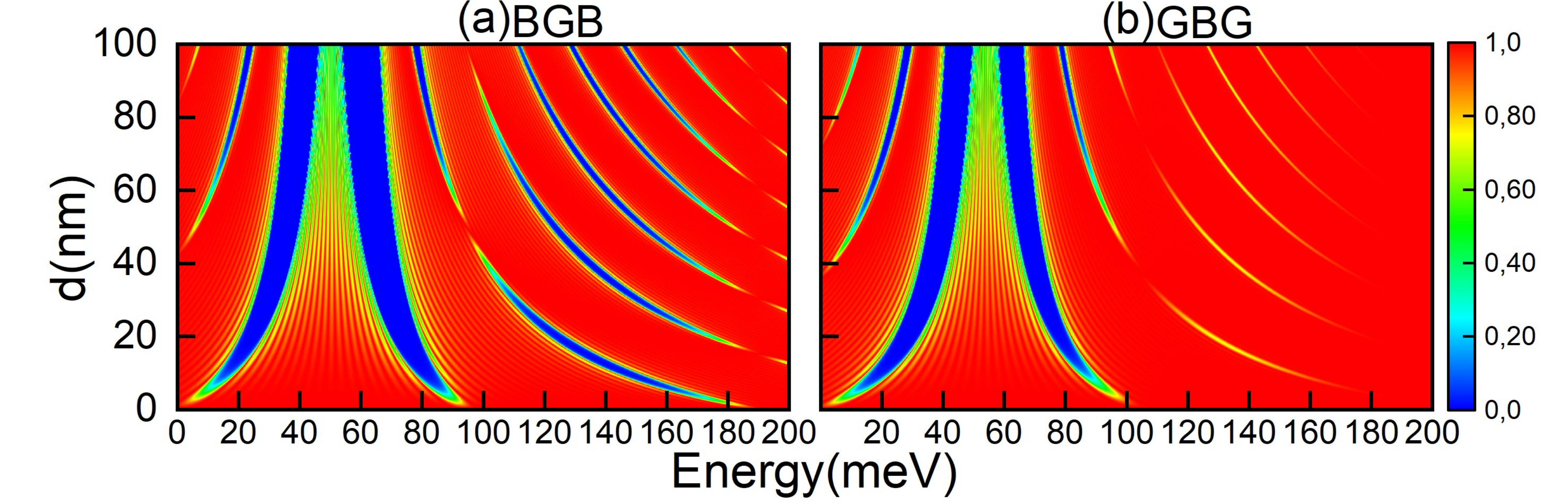}
		\caption{(Color online) The same as in Fig.~\ref{transEPhi} but now as a function of the energy and the barrier width $d$ for $w=20 $nm and $\theta=10^\circ$.}\label{transEd}
	\end{figure}
	%%%%%%%%%%%%%%
	%%%%%%%%%%%%%%%
	\begin{figure}[t!]
		\centering
		\includegraphics[width=\linewidth]{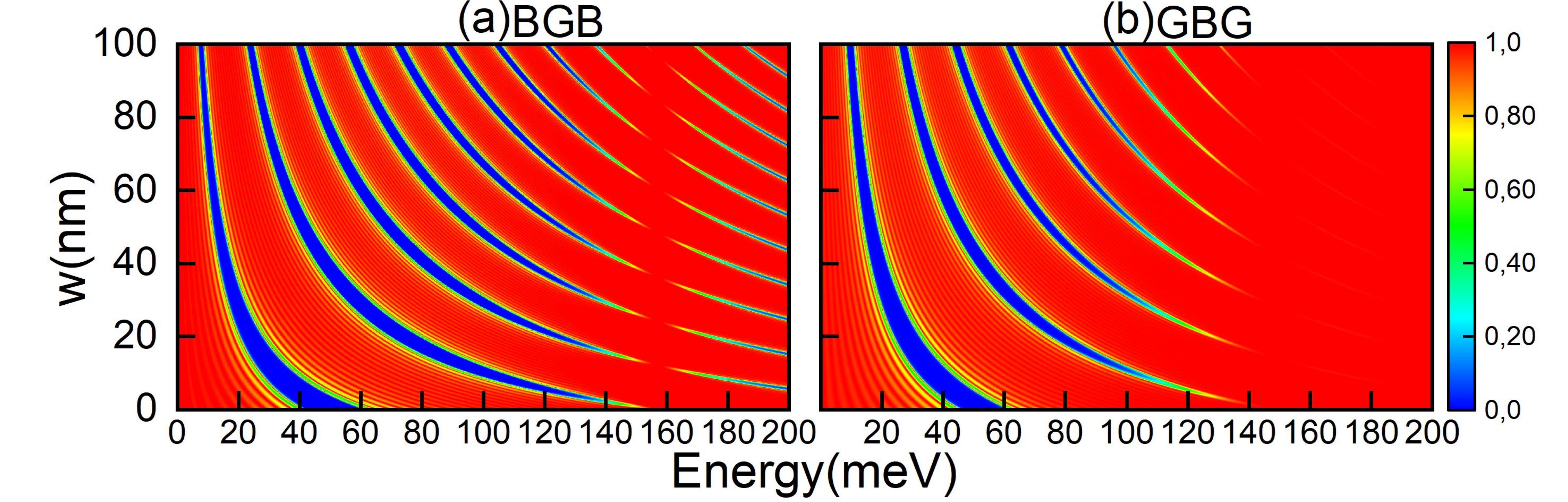}
		\caption{(Color online) The same as in Fig.~\ref{transEPhi} but now as a function of the energy and the well width $w$ for $d=20$ nm and $\theta=10^\circ$.}\label{transEw}
	\end{figure}
	%%%%%%%%%%%%%%
	%%%%%%%%%%%%%%%
	\begin{figure}[t!]
		\centering
		\includegraphics[width=\linewidth]{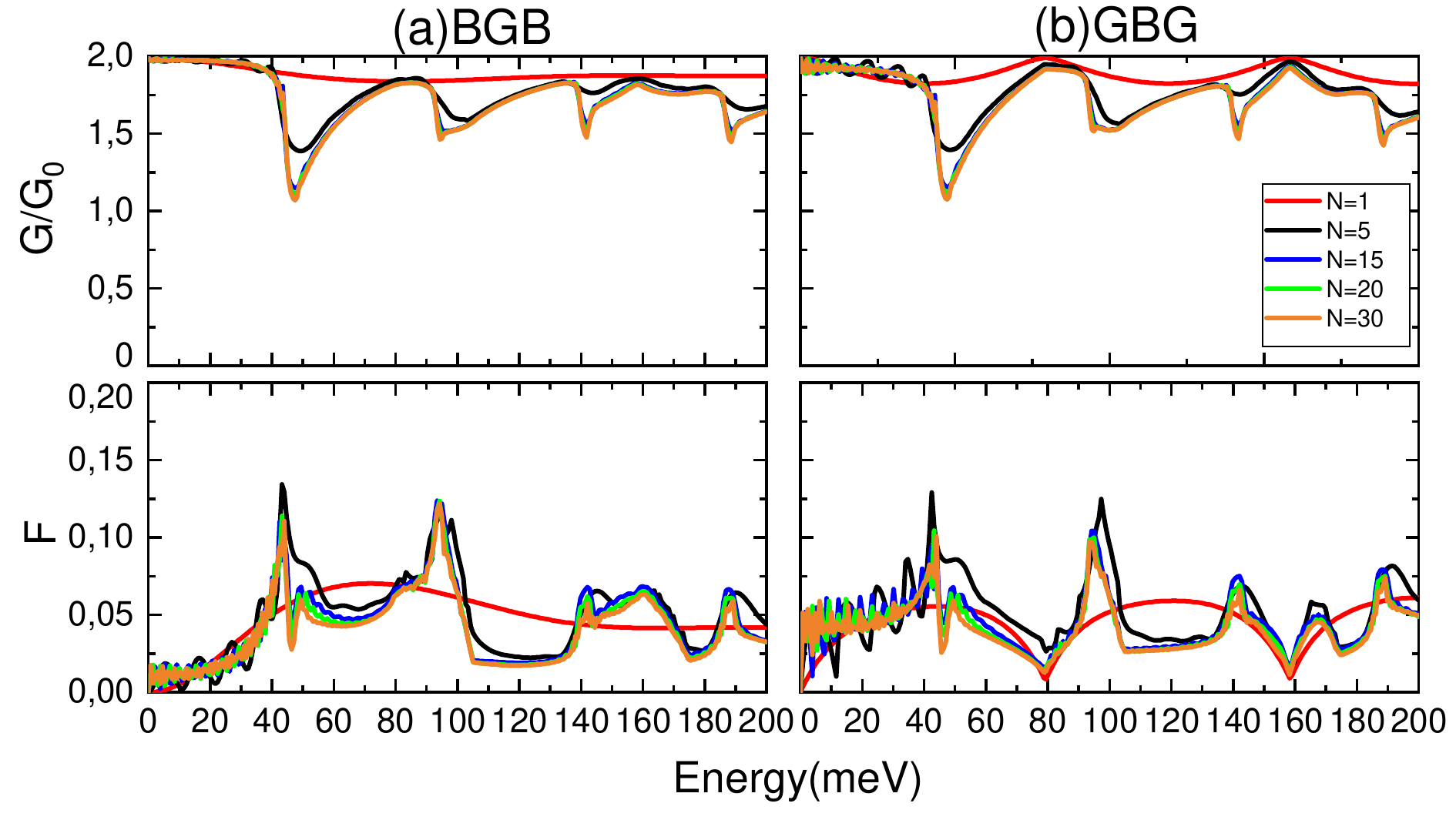}
		\caption{(Color online) Conductance and Fano factor as a function of the energy for BGB (a) and GBG (b), with $V_0=0$ meV and  $d=w=20$ nm.}\label{Conductance-and-Fano-factor-V=0}
	\end{figure}
	%%%%%%%%%%%%%%
	%%%%%%%%%%%%%%
	\begin{figure}[t!]
		\centering
		\includegraphics[width=\linewidth]{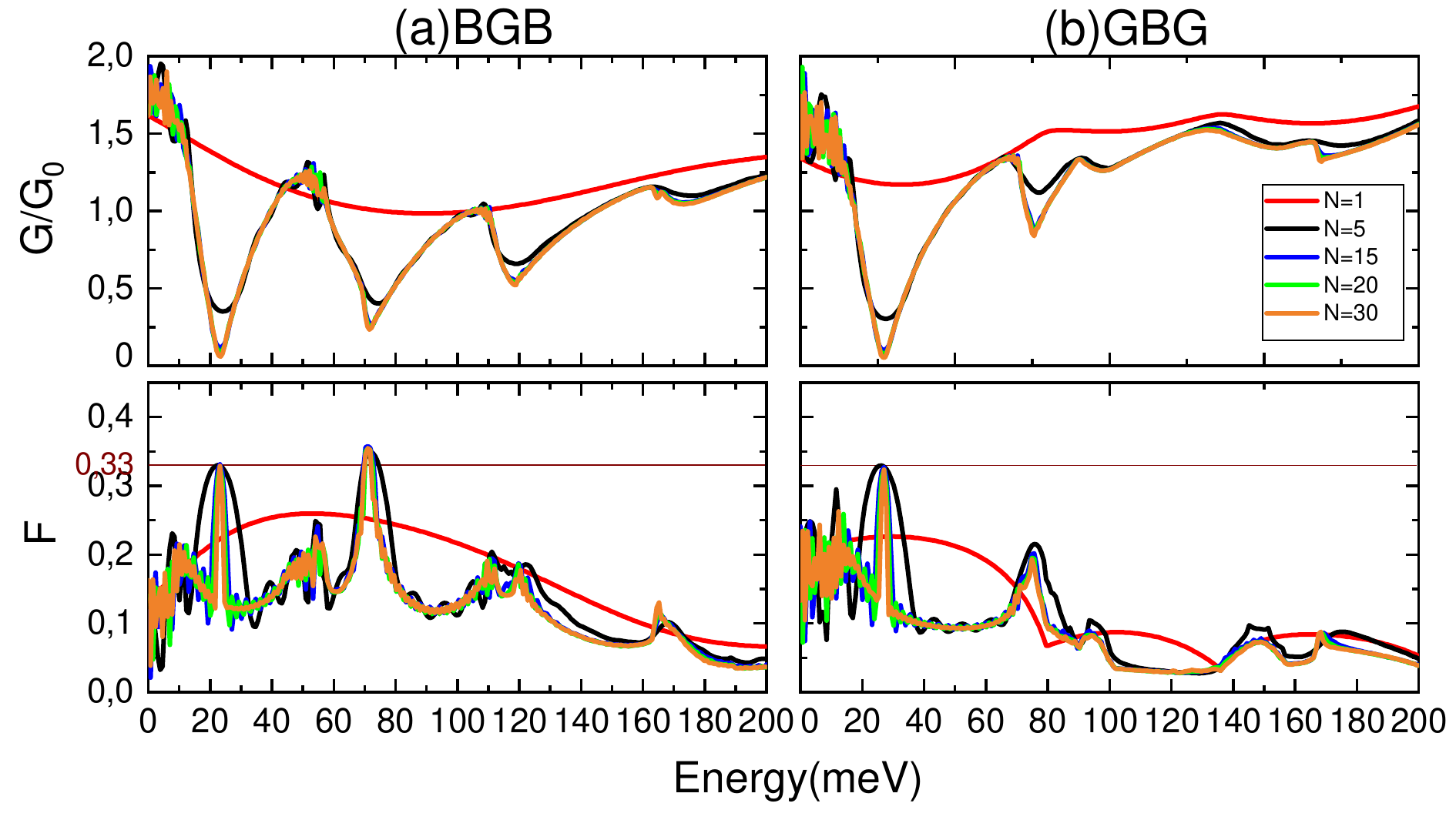}
		\caption{(Color online)  The same as in Fig.~\ref{Conductance-and-Fano-factor-V=0} but now for $V_0=50$ meV.}\label{Conductance-and-fano-factor-v=50}
	\end{figure}
	%%%%%%%%%%%%%%
	
	In Fig.~\ref{transEV}, we show the transmission probability
	as a function of the energy and barrier height for $\theta=10^\circ$, with the same conditions as in   Fig.~\ref{transEPhi}. We observe that with an increase in \( V_0 \), certain transmission gaps emerge. Remarkably, these gaps display dynamic behavior, with their centers gradually shifting to the right and their widths expanding proportionally. Notably, at \(V_0 = 120\) meV, a transmission gap emerges within the low-energy spectrum. As \(V_0\) increases, this gap expands into a transmission gap, which is a result that has been discussed in \cite{XU2015188superlattices}.

	Fig.~\ref{transEd} illustrates how the transmission probability varies with respect to the barrier width $d$ for BGB (a) and GBG (b) junctions across multiple barriers. As $d$ increases, we observe two distinct transmission gaps whose widths progressively increase {with increasing the barrier width $d$}. Additionally, Fig.~\ref{transEd} demonstrates the appearance and disappearance of transmission gaps across different energy regions as $d$ increases.
	
	Fig.~\ref{transEw} shows the transmission probability as a function of well width $w$. At $E=50$ meV, for $w=0$, we find the emergence of a transmission gap corresponding to a single barrier. {Remarkably, as $w$ increases}, this gap gets smaller while other transmission gaps arise. As $w$ rises further, these gaps become transmission gaps, first growing and then shrinking again. This phenomenon raises the prospect of using well width increases to create electron wave channels that are thinner \cite{XU2015188superlattices}. Moreover, 
	Fig.~\ref{transEw} illustrates how the center of each transmission gap gradually shifts to the left as $d$ rises, moving neighboring gaps closer together.

	Fig.~\ref{Conductance-and-Fano-factor-V=0} displays the conductance and Fano factor as a function of the energy for BGB (a) and GBG (b) junctions, with \( V_0 = 0 \) and \( n = 1, 5, 15, 20, 30 \). As expected, a noteworthy observation emerges: the minima in conductance align with the maxima in the Fano factor. Additionally, modifying the number of cells or the junction showcases the ability to modulate both the amplitude and the position of certain peaks in the Fano factor.

	Fig.~\ref{Conductance-and-fano-factor-v=50} presents the conductance and Fano factor for $V_0=50$ meV and with the same conditions as in Fig.~\ref{Conductance-and-Fano-factor-V=0}. We note that the presence of a potential barrier between layers introduces notable changes in the behavior of both the conductance and Fano factor. The precise location is $E=V/2$ with a constant Fermi velocity \cite{lima2015electronic}. We observe that, independently of $n$, the minimum conductance occurs near the Dirac point, corresponding to a Fano factor  $F=0.33$. This is a robust property in both graphene and borophene. These findings coincide with the results of a study conducted \cite{lima2018tuning} on the impact of Fermi velocity modulation on the Fano factor of graphene superlattices. In this case, when $V_0$ is introduced, the conductance decreases but the Fano factor increases. Interestingly, although the maximum value of the Fano factor tends to grow with the introduction of $V_0$, the amplitude of this increase can be decreased by carefully regulating the number of barriers in BGB and GBG
	junctions.

	%%%%%%%%%%%%%%%
	\begin{figure}[tbh]
		\centering
		\includegraphics[width=\linewidth]{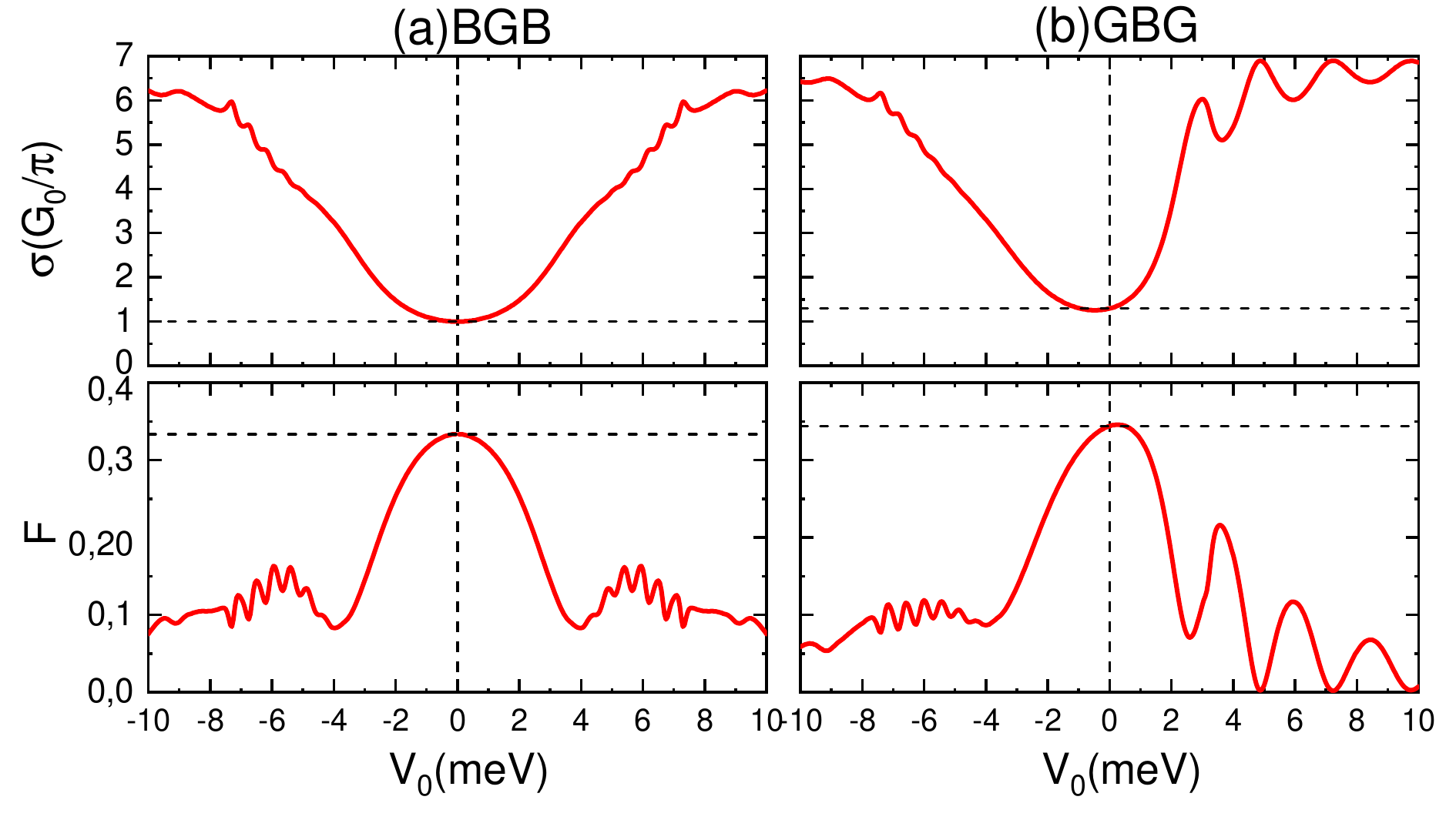}
		\caption{(Color online) Ballistic {scaled conductance} and Fano factor at the Dirac point as a function of the barrier height $V_0$ for BGB (a) and GBG (b), with $E=0$, $V_{\infty}=-\infty$ and  $D/d=5$.}\label{ConductanceBallisticV}
	\end{figure}
	%%%%%%%%%%%%%%%
	\begin{figure}[tbh]
		\centering
		\includegraphics[width=\linewidth]{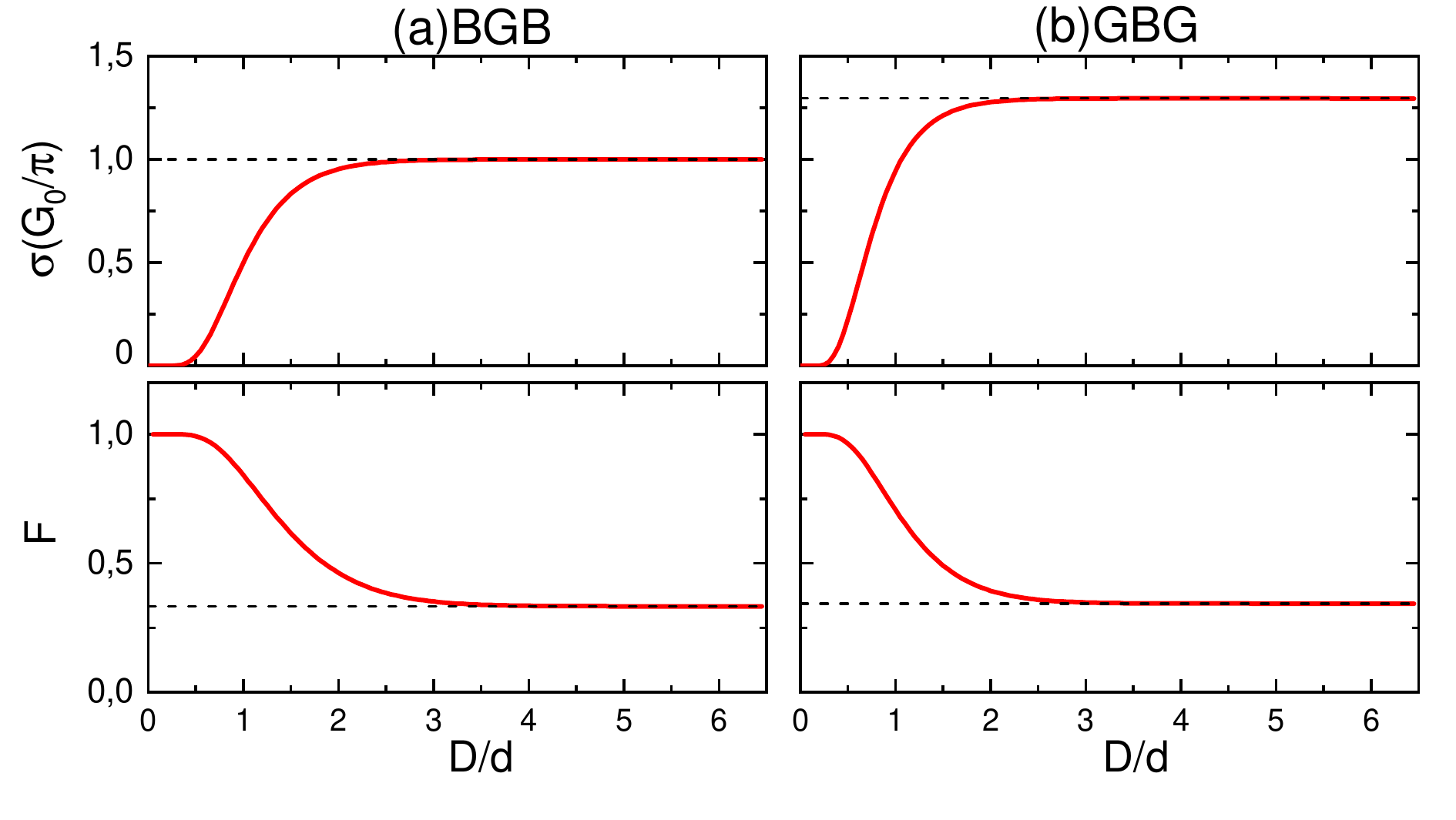}
		\caption{(Color online) The same as in 
			Fig.~\ref{ConductanceBallisticV} but now as a function of the width over length ratio $D/d$, with $E=V_0=0$ and $V_{\infty}=-\infty$.}\label{ConductanceBallisticD}
	\end{figure}
	%%%%%%%%%%%%%%
	
	At this level, we discuss the effect of tilted energy on {scaled conductance} $\sigma=Gd/D$ and Fano factor $F$ in the ballistic regime. Such ballistic transport requires heavy doping of regions 1 and 3 with the electrostatic potential $V(x)=V_{\infty}$ for $x<0$ and $x>d$ and $V(x)=V_{0}$ for $0<x<d$. In Fig. \ref{ConductanceBallisticV}, we present the ballistic {scaled conductance} and and Fano factor as functions of the barrier height $V_0$. 
	At the Dirac point ($E=V_0=0$) and for $V_{\infty}=-\infty$, for the BGB junction in Fig.~\ref{ConductanceBallisticV} (a), we observe the universal minimum ballistic {scaled conductance} $G_0/\pi$ and the Fano factor $F=0.333$, similarly to SLG \cite{Beenakker2006SubPoissonian}. Conversely, for the GBG junction in Fig. \ref{ConductanceBallisticV} (b), we have the results $1.295G_0/\pi$ and $F=0.343$. Notably, in GBG junctions, the {scaled conductance} is larger and the Fano factor smaller for $V_0 < 0$ compared to $V_0 > 0$. Additionally, it is noteworthy that BGB junction exhibits a symmetry at the Dirac point, in contrast to GBG junction.

	Fig.~\ref{ConductanceBallisticD} shows the ballistic {scaled conductance} and Fano factor  as a function of the width over length ratio $D/d$. For the BGB junction in Fig.~\ref{ConductanceBallisticD} (a), at the Dirac point ($E=V_0=0$), an interesting behavior is noticed as the limiting case $D/d\rightarrow\infty$.
	Specifically, we note that the ballistic {scaled conductance} \(\sigma\) tends towards \(G_0/\pi\) and the Fano factor \(F\) converges to \(0.333\), as reported in \cite{Beenakker2006SubPoissonian}. Conversely, for the GBG junction in Fig.~\ref{ConductanceBallisticD} (b), we see that \(\sigma\) approaches \(1.295G_0/\pi\), and the \(F\) converges to \(0.343\) when \(D/d\rightarrow\infty\). These results shed light on the relation between the length ratio of the junction and its ballistic transport, offering insights into the underlying electronic properties of the BGB and GBG junctions.

	%%%%%%%%%%%%%%%%%%%%%%%%%%%%%%%%%%%%%%%%%%%%%%%%%%
	\section{conclusion}
	%%%%%%%%%%%%%%%%%%%%%%%%%%%%%%%%%%%%%%%%%%%%%%%%%%
	\label{conclusions}
	%%%%%%%%%%%%%%%%%%%%%%%%%%%%%%%%%%%%%%%%%%%%%%%%%%%%%%%%%%
We have studied the transport properties of BGB and GBG junctions by separately treating the cases of a single barrier and multiple barriers in a superlattice configuration. In the framework of the first case, we have demonstrated that the two junctions have exceptional transport properties, particularly showing perfect transmission at normal incidence. We have seen that a decrease in the barrier height leads to a loss of transmission. By varying the energy and transverse wave vector, we observed pronounced scattered transmission at normal incidence. In the case of multiple barriers, we found a strong superlattice transmission, with the BGB junction retaining perfect transparency even at non-zero incident angles. It turns out that by varying the incident energy, many gaps appear in the transmission probability, and by adjusting the number of cells, incident angle, and barrier parameters, one can modify the number, width, and position of these transmission gaps.
{Strong oscillations in conductance and Fano factor were observed under diffuse transport conditions, demonstrating the extreme sensitivity of BGB and GBG junctions to variations in physical parameters.}
Notably, distinct patterns emerged when comparing the behavior of BGB and GBG junctions, particularly with respect to their responses to changes in barrier height.
In ballistic transport, our findings suggest that the minimum {scaled conductance} coincides with a maximum Fano factor. 
 {We also observed considerable differences between BGB and GBG junctions,} especially for {scaled conductance} and the Fano factor, which are influenced by the variation of the barrier height. Furthermore, our study of the length ratio (geometrical factor) has revealed intriguing patterns, as {scaled conductance} and the Fano factor approach specific values as the ratio approaches infinity. These results could offer valuable insights for developing electronic junctions utilizing graphene and borophene materials.

%\begin{thebibliography}{99}
%\end{thebibliography}

	\bibliographystyle{unsrt}
	\bibliography{Reff}

\begin{thebibliography}{10}

\bibitem{novoselov2004electric}
S.~V. Morozov D. Jiang Y. Zhang S. V. Dubonos I. V.~Grigorieva K.~S.~Novoselov,
  A. K.~Geim and A.~A. Firsov.
\newblock Electric field effect in atomically thin carbon films.
\newblock {\em Science}, 306(5696):666--669, 2004.

\bibitem{Novoselov2005two}
S.~V. Morozov D. Jiang M. I. Katsnelson I. V. Grigorieva S. V.~Dubonos
  K.~S.~Novoselov, A. K.~Geim and A.~A. Firsov.
\newblock Two-dimensional gas of massless dirac fermions in graphene.
\newblock {\em Nature}, 438(7065):197--200, 2005.

\bibitem{neto2009electronic}
AH~Castro Neto, Francisco Guinea, Nuno~MR Peres, Kostya~S Novoselov, and
  Andre~K Geim.
\newblock The electronic properties of graphene.
\newblock {\em Reviews of modern physics}, 81(1):109, 2009.

\bibitem{silvestre2015folded}
Ive Silvestre, Arthur~W Barnard, Samantha~P Roberts, Paul~L McEuen, and
  Rodrigo~G Lacerda.
\newblock Folded graphene nanochannels via pulsed patterning of graphene.
\newblock {\em Applied Physics Letters}, 106(15), 2015.

\bibitem{zhao2015graphene}
Xiaoli Zhao, Bingna Zheng, Tieqi Huang, and Chao Gao.
\newblock Graphene-based single fiber supercapacitor with a coaxial structure.
\newblock {\em Nanoscale}, 7(21):9399--9404, 2015.

\bibitem{ji2012atomic}
Shuai-Hua Ji, JB~Hannon, RM~Tromp, V~Perebeinos, J~Tersoff, and FM~Ross.
\newblock Atomic-scale transport in epitaxial graphene.
\newblock {\em Nature materials}, 11(2):114--119, 2012.

\bibitem{piazza2014planar}
Wei-Li Li Ya-Fan Zhao Jun~Li Zachary A.~Piazza, Han-Shi~Hu and Lai-Sheng Wang.
\newblock Planar hexagonal $b_{36}$ as a potential basis for extended
  single-atom layer boron sheets.
\newblock {\em Nature communications}, 5(1):3113, 2014.

\bibitem{li20202d}
Dengfeng Li, Junfeng Gao, Peng Cheng, Jia He, Yan Yin, Yanxiao Hu, Lan Chen,
  Yuan Cheng, and Jijun Zhao.
\newblock 2d boron sheets: structure, growth, and electronic and thermal
  transport properties.
\newblock {\em Advanced Functional Materials}, 30(8):1904349, 2020.

\bibitem{kaneti2021borophene}
Yusuf~Valentino Kaneti, Didi~Prasetyo Benu, Xingtao Xu, Brian Yuliarto, Yusuke
  Yamauchi, and Dmitri Golberg.
\newblock Borophene: two-dimensional boron monolayer: synthesis, properties,
  and potential applications.
\newblock {\em Chemical Reviews}, 122(1):1000--1051, 2021.

\bibitem{zhang2017two}
Zhuhua Zhang, Evgeni~S Penev, and Boris~I Yakobson.
\newblock Two-dimensional boron: structures, properties and applications.
\newblock {\em Chemical Society Reviews}, 46(22):6746--6763, 2017.

\bibitem{Zhang2018Oblique}
Shu-Hui Zhang and Wen Yang.
\newblock Oblique klein tunneling in $8\ensuremath{-}pmmn$ borophene
  $p\ensuremath{-}n$ junctions.
\newblock {\em Phys. Rev. B}, 97:235440, 2018.

\bibitem{kong2021oblique}
Zhan Kong, Jian Li, Yi~Zhang, Shu-Hui Zhang, and Jia-Ji Zhu.
\newblock Oblique and asymmetric klein tunneling across smooth np junctions or
  npn junctions in 8-pmmn borophene.
\newblock {\em Nanomaterials}, 11(6):1462, 2021.

\bibitem{lopez2016electronic}
Alejandro Lopez-Bezanilla and Peter~B Littlewood.
\newblock Electronic properties of 8- pmmn borophene.
\newblock {\em Physical Review B}, 93(24):241405, 2016.

\bibitem{gonzalez2008boron}
Nevill Gonzalez~Szwacki.
\newblock Boron fullerenes: a first-principles study.
\newblock {\em Nanoscale Research Letters}, 3:49--54, 2008.

\bibitem{tang2007novel}
Hui Tang and Sohrab Ismail-Beigi.
\newblock Novel precursors for boron nanotubes: the competition of two-center
  and three-center bonding in boron sheets.
\newblock {\em Physical Rreview Letters}, 99(11):115501, 2007.

\bibitem{tang2009self}
Hui Tang and Sohrab Ismail-Beigi.
\newblock Self-doping in boron sheets from first principles: A route to
  structural design of metal boride nanostructures.
\newblock {\em Physical Review B}, 80(13):134113, 2009.

\bibitem{xu2016hydrogenated}
Li-Chun Xu, Aijun Du, and Liangzhi Kou.
\newblock Hydrogenated borophene as a stable two-dimensional dirac material
  with an ultrahigh fermi velocity.
\newblock {\em Physical Chemistry Chemical Physics}, 18(39):27284--27289, 2016.

\bibitem{jugovac2023coupling}
Matteo Jugovac, Iulia Cojocariu, Carlo~Alberto Brondin, Alessandro Crotti,
  Marin Petrovi{\'c}, Stefano Bonetti, Andrea Locatelli, and Tevfik~Onur
  Mente{\c{s}}.
\newblock Coupling borophene to graphene in air-stable heterostructures.
\newblock {\em Advanced Electronic Materials}, 9(8):2300136, 2023.

\bibitem{liu2019borophene}
Xiaolong Liu and Mark~C Hersam.
\newblock Borophene-graphene heterostructures.
\newblock {\em Science Advances}, 5(10):eaax6444, 2019.

\bibitem{hou2021borophene}
Chuang Hou, Guo’an Tai, Bo~Liu, Zenghui Wu, and Yonghe Yin.
\newblock Borophene-graphene heterostructure: Preparation and ultrasensitive
  humidity sensing.
\newblock {\em Nano Research}, pages 1--8, 2021.

\bibitem{wang2020activating}
Shipeng Wang, Qingsong Li, Kang Hu, Qiangchun Liu, Xiaofang Liu, and Xiangkai
  Kong.
\newblock Activating microwave absorption performance by reduced graphene
  oxide-borophene heterostructure.
\newblock {\em Composites Part A: Applied Science and Manufacturing},
  138:106033, 2020.

\bibitem{mortazavi2020machine}
Bohayra Mortazavi, Evgeny~V Podryabinkin, Stephan Roche, Timon Rabczuk,
  Xiaoying Zhuang, and Alexander~V Shapeev.
\newblock Machine-learning interatomic potentials enable first-principles
  multiscale modeling of lattice thermal conductivity in graphene/borophene
  heterostructures.
\newblock {\em Materials Horizons}, 7(9):2359--2367, 2020.

\bibitem{das2015beyond}
Saptarshi Das, Joshua~A Robinson, Madan Dubey, Humberto Terrones, and Mauricio
  Terrones.
\newblock Beyond graphene: progress in novel two-dimensional materials and van
  der waals solids.
\newblock {\em Annual Review of Materials Research}, 45:1--27, 2015.

\bibitem{butler2013progress}
Sheneve~Z Butler, Shawna~M Hollen, Linyou Cao, Yi~Cui, Jay~A Gupta, Humberto~R
  Guti{\'e}rrez, Tony~F Heinz, Seung~Sae Hong, Jiaxing Huang, Ariel~F Ismach,
  et~al.
\newblock Progress, challenges, and opportunities in two-dimensional materials
  beyond graphene.
\newblock {\em ACS Nano}, 7(4):2898--2926, 2013.

\bibitem{ferrari2015science}
Andrea~C Ferrari, Francesco Bonaccorso, Vladimir Fal'Ko, Konstantin~S
  Novoselov, Stephan Roche, Peter B{\o}ggild, Stefano Borini, Frank~HL Koppens,
  Vincenzo Palermo, Nicola Pugno, et~al.
\newblock Science and technology roadmap for graphene, related two-dimensional
  crystals, and hybrid systems.
\newblock {\em Nanoscale}, 7(11):4598--4810, 2015.

\bibitem{xie2021chemistry}
Zhongjian Xie, Bin Zhang, Yanqi Ge, Yao Zhu, Guohui Nie, YuFeng Song,
  Chang-Keun Lim, Han Zhang, and Paras~N Prasad.
\newblock Chemistry, functionalization, and applications of recent
  monoelemental two-dimensional materials and their heterostructures.
\newblock {\em Chemical Reviews}, 122(1):1127--1207, 2021.

\bibitem{tan2017recent}
Chaoliang Tan, Xiehong Cao, Xue-Jun Wu, Qiyuan He, Jian Yang, Xiao Zhang, Junze
  Chen, Wei Zhao, Shikui Han, Gwang-Hyeon Nam, et~al.
\newblock Recent advances in ultrathin two-dimensional nanomaterials.
\newblock {\em Chemical Reviews}, 117(9):6225--6331, 2017.

\bibitem{ma2020review}
Rong Ma, Jie Sun, Dong~Hui Li, and Jin~Jia Wei.
\newblock Review of synergistic photo-thermo-catalysis: Mechanisms, materials
  and applications.
\newblock {\em International Journal of Hydrogen Energy}, 45(55):30288--30324,
  2020.

\bibitem{usman2021bismuth}
Muhammad Usman, Muhammad Humayun, Syed~Shaheen Shah, Habib Ullah, Asif~A Tahir,
  Abbas Khan, and Habib Ullah.
\newblock Bismuth-graphene nanohybrids: synthesis, reaction mechanisms, and
  photocatalytic applications—a review.
\newblock {\em Energies}, 14(8):2281, 2021.

\bibitem{riazi2021ti}
Hossein Riazi, Srinivasa~Kartik Nemani, Michael~C Grady, Babak Anasori, and
  Masoud Soroush.
\newblock Ti 3 c 2 mxene--polymer nanocomposites and their applications.
\newblock {\em Journal of Materials Chemistry A}, 9(13):8051--8098, 2021.

\bibitem{iqbal2023nanostructures}
Muhammad~Aamir Iqbal, Nadia Anwar, Maria Malik, Mohammed Al-Bahrani, Md~Rasidul
  Islam, Jeong~Ryeol Choi, Phuong~V Pham, and Xiaofeng Liu.
\newblock Nanostructures/graphene/silicon junction-based high-performance
  photodetection systems: progress, challenges, and future trends.
\newblock {\em Advanced Materials Interfaces}, 10(7):2202208, 2023.

\bibitem{abdullah2017quantum}
Hasan~M Abdullah, B~Van~Duppen, M~Zarenia, Hocine Bahlouli, and FM~Peeters.
\newblock Quantum transport across van der waals domain walls in bilayer
  graphene.
\newblock {\em Journal of Physics: Condensed Matter}, 29(42):425303, 2017.

\bibitem{PhysRevB.108.245419}
Nadia Benlakhouy, Ahmed Jellal, and Michael Schreiber.
\newblock Transport properties of hybrid single-bilayer graphene interfaces in
  a magnetic field.
\newblock {\em Physical Review B}, 108(10):245419, 2023.

\bibitem{behura2019graphene}
Sanjay~K Behura, Chen Wang, Yu~Wen, and Vikas Berry.
\newblock Graphene--semiconductor heterojunction sheds light on emerging
  photovoltaics.
\newblock {\em Nature Photonics}, 13(5):312--318, 2019.

\bibitem{khalatbari2021spin}
H~Khalatbari, S~Izadi Vishkayi, and H~Rahimpour Soleimani.
\newblock Spin transport properties in tm-doped b38 fullerene/borophene
  junctions.
\newblock {\em Physica B: Condensed Matter}, 621:413284, 2021.

\bibitem{xu2023valley}
Yafang Xu, Yu~Fang, and Guojun Jin.
\newblock Valley-polarized and supercollimated electronic transport in an
  8-pmmn borophene superlattice.
\newblock {\em New Journal of Physics}, 25(1):013020, 2023.

\bibitem{feng2017dirac}
Baojie Feng, Osamu Sugino, Ro-Ya Liu, Jin Zhang, Ryu Yukawa, Mitsuaki Kawamura,
  Takushi Iimori, Howon Kim, Yukio Hasegawa, Hui Li, et~al.
\newblock Dirac fermions in borophene.
\newblock {\em Physical review letters}, 118(9):096401, 2017.

\bibitem{sadhukhan2017anisotropic}
Krishanu Sadhukhan and Amit Agarwal.
\newblock Anisotropic plasmons, friedel oscillations, and screening in 8-$pmmn$
  borophene.
\newblock {\em Physical Review B}, 96(3):035410, 2017.

\bibitem{nakhaee2018tight}
Mohammad Nakhaee, SA~Ketabi, and FM~Peeters.
\newblock Tight-binding model for borophene and borophane.
\newblock {\em Physical Review B}, 97(12):125424, 2018.

\bibitem{Zhou2019Valley}
Xingfei Zhou.
\newblock Valley-dependent electron retroreflection and anomalous klein
  tunneling in an 8-$pmmn$ borophene-based
  $n\text{\ensuremath{-}}p\text{\ensuremath{-}}n$ junction.
\newblock {\em Phys. Rev. B}, 100:195139, 2019.

\bibitem{zabolotskiy2016strain}
AD~Zabolotskiy and Yu~E Lozovik.
\newblock Strain-induced pseudomagnetic field in the dirac semimetal borophene.
\newblock {\em Physical Review B}, 94(16):165403, 2016.

\bibitem{katsnelson2006chiral}
Mikhail~Iosifovich Katsnelson, Konstantin~Sergejevi{\v{c}} Novoselov, and
  Andre~Konstantin Geim.
\newblock Chiral tunnelling and the klein paradox in graphene.
\newblock {\em Nature physics}, 2(9):620--625, 2006.

\bibitem{wang2010electronic}
Li-Gang Wang and Shi-Yao Zhu.
\newblock Electronic band gaps and transport properties in graphene
  superlattices with one-dimensional periodic potentials of square barriers.
\newblock {\em Physical Review B}, 81(20):205444, 2010.

\bibitem{wang2014transfer}
Yu~Wang.
\newblock Transfer matrix theory of monolayer graphene/bilayer graphene
  heterostructure superlattice.
\newblock {\em Journal of Applied Physics}, 116(16), 2014.

\bibitem{li2009generalized}
Haidong Li, Lin Wang, Zhihuan Lan, and Yisong Zheng.
\newblock Generalized transfer matrix theory of electronic transport through a
  graphene waveguide.
\newblock {\em Physical Review B}, 79(15):155429, 2009.

\bibitem{XU2015188superlattices}
Yi~Xu, Ying He, and Yanfang Yang.
\newblock Transmission gaps in graphene superlattices with periodic potential
  patterns.
\newblock {\em Physica B: Condensed Matter}, 457:188--193, 2015.

\bibitem{mason2002chebyshev}
John~C Mason and David~C Handscomb.
\newblock {\em Chebyshev polynomials}.
\newblock CRC press, 2002.

\bibitem{covaci2010efficient}
L~Covaci, FM~Peeters, and M~Berciu.
\newblock Efficient numerical approach to inhomogeneous superconductivity: the
  chebyshev-bogoliubov--de gennes method.
\newblock {\em Physical review letters}, 105(16):167006, 2010.

\bibitem{buttiker1985generalized}
M~B{\"u}ttiker, Y~Imry, R~Landauer, and S~Pinhas.
\newblock Generalized many-channel conductance formula with application to
  small rings.
\newblock {\em Physical Review B}, 31(10):6207, 1985.

\bibitem{Beenakker2006SubPoissonian}
J.~Tworzyd\l{}o, B.~Trauzettel, M.~Titov, A.~Rycerz, and C.~W.~J. Beenakker.
\newblock Sub-poissonian shot noise in graphene.
\newblock {\em Phys. Rev. Lett.}, 96(4):246802, 2006.

\bibitem{lima2018tuning}
Jonas~RF Lima, Anderson~LR Barbosa, Claudionor~Gomes Bezerra, and Luiz Felipe~C
  Pereira.
\newblock Tuning the fano factor of graphene via fermi velocity modulation.
\newblock {\em Physica E: Low-dimensional Systems and Nanostructures},
  97:105--110, 2018.

\bibitem{van2013four}
B~Van~Duppen and FM~Peeters.
\newblock Four-band tunneling in bilayer graphene.
\newblock {\em Physical Review B}, 87(20):205427, 2013.

\bibitem{PhysRevLett.132.056204}
Tianlin Li, Hanying Chen, Kun Wang, Yifei Hao, Le~Zhang, Kenji Watanabe,
  Takashi Taniguchi, and Xia Hong.
\newblock Transport anisotropy in one-dimensional graphene superlattice in the
  high kronig-penney potential limit.
\newblock {\em Phys. Rev. Lett.}, 132:056204, Jan 2024.

\bibitem{dakhlaoui2021quantum}
Hassen Dakhlaoui, Walid Belhadj, and Bryan~M Wong.
\newblock Quantum tunneling mechanisms in monolayer graphene modulated by
  multiple electrostatic barriers.
\newblock {\em Results in Physics}, 26:104403, 2021.

\bibitem{dakhlaoui2021modulating}
Hassen Dakhlaoui, Shaffa Almansour, Walid Belhadj, and Bryan~M Wong.
\newblock Modulating the conductance in graphene nanoribbons with
  multi-barriers under an applied voltage.
\newblock {\em Results in Physics}, 27:104505, 2021.

\bibitem{bai2007klein}
Chunxu Bai and Xiangdong Zhang.
\newblock Klein paradox and resonant tunneling in a graphene superlattice.
\newblock {\em Physical Review B}, 76(7):075430, 2007.

\bibitem{lima2015electronic}
Jonas~RF Lima.
\newblock Electronic structure of a graphene superlattice with massive dirac
  fermions.
\newblock {\em Journal of Applied Physics}, 117(8), 2015.

\end{thebibliography}
\end{document}